\documentclass[journal]{IEEEtran}

\usepackage{cite}
\usepackage{amsmath,amssymb,amsfonts, amsthm}
\usepackage{acronym}
\usepackage{graphicx}
\usepackage{textcomp}
\usepackage{xcolor}
\usepackage{threeparttable}
\usepackage{booktabs}

\usepackage{pifont}
\usepackage[font=small]{caption}
\usepackage{subcaption}
\usepackage[hidelinks]{hyperref}
\usepackage{comment}
\usepackage{upgreek}

\usepackage{multirow}

\newcommand{\bfb}[1]{\boldsymbol{\rm #1}}
\newcommand{\bfg}[1]{\boldsymbol{#1}}
\newcommand{\nx}{\nu}
\newcommand{\ny}{\mu}
\newcommand{\nd}{r}

\newcommand{\xs}{\ensuremath{\mbox{$\bfb {x}$}}}

\newcommand{\jac}[2]{\bfg{{#1}}_{\hspace{-0.2mm}#2}}

\newcommand{\T}{^{\intercal}}
\newcommand{\AS}{\bfb A}

\newcommand{\Etdi}{\bfb{F}}
\newcommand{\Atdi}{\bfb{G}}

\acrodef{rocof}[RoCoF]{rate of change of frequency}
\acrodef{hm}[HM]{Heun's method}
\acrodef{fem}[FEM]{forward Euler method}
\acrodef{tm}[TM]{trapezoidal method}
\acrodef{pf}[PF]{participation factor}
\acrodef{sssa}[SSSA]{small-signal stability analysis}
\acrodef{bem}[BEM]{backward Euler method}
\acrodef{dae}[DAE]{differential algebraic equation}
\acrodef{ddae}[DDAE]{delay differential algebraic equation}
\acrodef{pll}[PLL]{phase-locked loop}
\acrodef{sm}[SM]{synchronous machine}
\acrodef{der}[DER]{distributed energy resource}

\acrodef{psa}[PSA]{partitioned-solution approach}
\acrodef{wams}[WAMS]{wide-area measurement system}

\acrodef{ode}[ODE]{ordinary differential equation}
\acrodef{tdi}[TDI]{time-domain integration}
\acrodef{agc}[AGC]{automatic generation control}
\acrodef{pss}[PSS]{power system stabilizer}
\acrodef{avr}[AVR]{automatic voltage regulator}
\acrodef{tg}[TG]{turbine governor}
\acrodef{srf-pll}[SRF-PLL]{synchronous reference frame phase-locked loop}
\acrodef{gte}[GTE]{global truncation error}

\definecolor{ethblau}{RGB}{33,92,175}

\begin{document}

\title{Stability of the Theta Method for Systems with Multiple Time-Delayed Variables

\author{Andreas Bouterakos, 
Georgios Tzounas, {\em IEEE Member}
  \thanks{The authors are with 
  the School of Electrical and Electronic Engineering, University College Dublin, Dublin, D04V1W8, Ireland.
  E-mails: 
  andreas.bouterakos@ucdconnect.ie,
  georgios.tzounas@ucd.ie
\textit{(Corresponding author:
Georgios Tzounas.)}
}
  } 
  }

\maketitle 

\begin{abstract} 
The paper focuses on the numerical stability and accuracy of implicit \ac{tdi} methods when applied for the solution of a power system model impacted by time delays. 
Such a model is generally formulated as a set of \acp{ddae} in non index-1 Hessenberg form.  In particular, the paper shows that numerically stable \ac{ode} methods, such as the trapezoidal and the Theta method, 
can become unstable when applied to a power system that includes a significant number of delayed variables.  Numerical stability is discussed through a scalar test delay differential equation, as well as through a matrix pencil approach that accounts for the \acp{ddae} of any given dynamic power system model.  Simulation results are presented in a case study based on the IEEE 39-bus system.  
\end{abstract}

\begin{IEEEkeywords}
Time-domain simulation, 
time delays,
Theta method,
numerical stability,
matrix pencils.
\end{IEEEkeywords}

\IEEEpeerreviewmaketitle

\section{Introduction}
\label{sec:intro}

\subsection{Motivation}

The most successful to date approach to study the dynamic response and performance of a power system following a large disturbance is to perform nonlinear time-domain simulations.  The focus of this paper is on how time-domain simulations are impacted in the presence of time delays, 
in particular how the numerical stability and accuracy of state-of-the-art \acf{tdi} methods can be challenged in the presence of multiple time-delayed variables.  

The topic is becoming more and more relevant with the increasing number of inverter-based \acp{der}, which gives rise to more points of control and coordination, and thus to more data that need to be processed and transferred.  As a result, the total number of control signals in the system impacted by delays is also increasing significantly.  In many cases, these delays cannot be neglected, as they can pose a threat to the overall system security, e.g.,~see 
\cite{motiv_1, motiv_2, motiv_3, liu2018stability, wadc_mul_del, Ledva_delays, Jin_delays}.


\subsection{Literature Review}

Power system models for short-term stability analysis are conventionally described by a set of nonlinear \acp{dae} \cite{kundur:94}.
These equations are known to be \textit{stiff}, as their time constants span multiple time scales.  To deal with system stiffness and ensure numerical stability, an implicit \acf{ode} method, such as the trapezoidal or the Theta method \cite{Tripathy1977,1995paserba, powerfactory},
is typically employed when conducting a \ac{tdi} of a power system model.  

Inclusion of time delays leads to the reformulation of the power system model into a set of \acfp{ddae}.  A common approach to handle these delays in \ac{tdi} is through modification of standard \ac{ode} numerical methods \cite{Milano2012889, Hiskens_1205019}. 
Nevertheless, the numerical stability properties of \ac{ode} methods do not transfer to time-delayed systems.  For example, it is known that no A-stable natural Runge–Kutta method is stable on the whole class of stable linear systems of delay differential equations \cite{bellen:03}.  Alternatively, more sophisticated, specialized methods for delay systems can be employed at an additional computational cost, e.g.,~we cite the family of Radau~IIA methods \cite{Guglielmi20011}. 

Selecting a proper \ac{tdi} method is generally a matter of finding a good trade-off between stability, precision and computational speed.  
In this context, the classical approach to test the stability properties of a \ac{tdi} method is by studying its response when applied to a linear scalar equation representing the class to which the examined system belongs.  This approach neglects the dynamics of the specific model examined and thus is not suitable for precision assessment.
Precision is typically estimated through truncation error analysis \cite{hairer:91:dae}.  
However, such an analysis is insufficient to 
predict numerical instabilities.
Aiming to address the limitations of these approaches, 
the second author of this paper has recently proposed a matrix pencil approach that allows studying stability and precision of \ac{tdi} methods in a unified way.
Such an approach was first formulated for \ac{dae} models in \cite{tzounas2022tdistab}, where numerical errors between power system dynamics and the dynamics of the discrete-time system that arises due to the application of the \ac{tdi} method were investigated through a comparison of their associated matrix pencils.  A further investigation was carried out in \cite{Mode-shape_def}, where participation factor analysis was used to assess the numerical deformation that standard \ac{tdi} methods introduce to the shape of the coupling between dynamic modes and system variables.  

The focus of the present work is on the numerical stability and accuracy of standard implicit \ac{tdi} methods when employed for the numerical solution of \ac{ddae} power system models.  In this vein, a preliminary study using the matrix pencil approach was carried out in 
\cite{TZOUNAS2022108266}.  Therein, results indicate that, including a delay that is 
multiple of the time step in a single variable does not have a notable effect on numerical stability.  
In this paper, we build upon previous work to provide new theoretical insights and studies suggesting that the number of time-delayed variables and the magnitudes of their coefficients in the \acp{ddae} are crucial factors impacting both numerical stability and accuracy and can lead otherwise very robust implicit methods to behave poorly.

\subsection{Contributions}

The paper provides -- to the best of our knowledge -- for the first time a comprehensive discussion  on how standard implicit \ac{tdi} methods, such as the trapezoidal and Theta method, can be destabilized when applied to power system models impacted by multiple time-delayed variables.  A proof of concept is first provided through a linear test delay differential equation.  Then, a systematic analysis that accounts for the dynamics of real-world power system models is carried out through proper extension of a matrix pencil approach recently proposed by the second author.  The potential of compensating numerical instabilities through simple adjustments of the damping parameter of the Theta method is also duly discussed.

\subsection{Paper Organization}

The remainder of the paper is organized as follows.  Section~\ref{sec:model} recalls the formulation and numerical integration of \ac{ddae} power system models.  The proposed numerical stability analysis is presented in Section~\ref{sec:num_stab}.  Section~\ref{sec:case} discusses the case study.  The case study first considers the standard IEEE 39-bus system and then a modified version of the same system that includes inverter-based \acp{der}.  Conclusions are drawn in Section~\ref{sec:conclusion}.

\section{Numerical Integration in the Presence of Delays}
\label{sec:model}

\subsection{Conventional DAE Model}

Power system dynamics can be described with a set of nonlinear \acp{dae}. 
Using an implicit  form:

\begin{align}
\label{eq:implicit}
\bfg 0_{\nx+\ny,1} &= \bfg \phi ({\bfg x}', \bfg x, \bfg y) \, .
\end{align}
In \eqref{eq:implicit}, $\bfg x = \bfg x(t) : [0,\infty) \rightarrow \mathbb{R}^{\nx}$ and $\bfg y = \bfg y(t): [0,\infty) \rightarrow \mathbb{R}^{\ny}$ are the state and algebraic variables, respectively, of the system; 
$\bfg \phi : \mathbb{R}^{\nx+\ny} \rightarrow \mathbb{R}^{\nx+\ny}$ are
nonlinear functions;
$\bfg 0_{\nx+\ny,1}$ denotes the zero matrix of dimensions $(\nx+\ny)\times 1$.  
Differential and algebraic equations are commonly expressed as two distinct sets, i.e., using the Hessenberg form \cite{kundur:94}:  
\begin{equation}
  \begin{aligned}
    \label{eq:dae}
{\bfg x}' &= \bfg f(\bfg x, \bfg y) \, , \\
\bfg 0_{\ny,1} &=  \bfg g(\bfg x, \bfg y  ) \, ,
  \end{aligned}    
\end{equation}
where $\bfg f : \mathbb{R}^{\nx+\ny} \rightarrow \mathbb{R}^{\nx}$ and $\bfg g : \mathbb{R}^{\nx+\ny} \rightarrow \mathbb{R}^{\ny}$ define, respectively, the system's differential and algebraic equations. 

Approximating the solution of \eqref{eq:dae} for a set of known initial conditions requires performing a time-domain simulation through the application of a proper discrete \ac{tdi} method.  Implicit \ac{ode} methods are commonly preferred from explicit ones due to their ability to handle system stiffness, as well as due to their numerical stability properties.  We recall here that the stability properties of implicit \ac{ode} methods are not guaranteed to hold when applied to \acp{dae}, but rather transfer better to \acp{dae} with a small differentiation index \cite{dif_index_1, März_1992, dif_index_4}.\footnote{The differentiation index $q$ of a \ac{dae} set in the form of \eqref{eq:dae} is defined as the highest order of time derivative $d^{q}/dt^{q}$ required to to eliminate the algebraic constraints.}
We note that \acp{dae} \eqref{eq:dae} are index-1 if 
$\bfg (\bfg x,\bfg y) := [\bfg x\T, \bfg y\T]\T$ 
(where $\T$ is the matrix transpose)
is differentiable and $\bfg g_y = \partial\bfg g/\partial\bfg y$ is not singular at every $t$ along the solution flow.

In this paper, we employ the well-known Theta method for the \ac{tdi} of \eqref{eq:dae}.  In this case, discretization is implemented through the following linear fractional transformation: 
\begin{equation}
\label{eq:theta:ztos}
    z = \frac{1+h\theta s}{1-h(1-\theta) s} 
    \Leftrightarrow
    s =  \frac{1}{h}
    \frac{z - 1}{(1-\theta) z+\theta } \, ,
\end{equation}
where $s$ and $z$ are the complex variables of the \textit{s}-domain and \textit{z}-domain, respectively; $h$ is the integration time step size; and $\theta$ defines the method’s damping. 
Applied to \eqref{eq:dae}, the Theta method reads as follows:
\begin{equation}
\begin{aligned}\label{eq:theta}
\bfg x_{n+1} &= \bfg x_{n} + h
[\theta\bfg f(\bfg x_{n},\bfg y_{n}) + 
(1-\theta)\bfg f(\bfg x_{n+1},\bfg y_{n+1})] \, , \\
\bfg 0_{\ny,1} 
&= h \bfg g(\bfg x_{n+1},\bfg y_{n+1}) \, .
\end{aligned}
\end{equation}
For the sake of completeness, the derivation of \eqref{eq:theta} from \eqref{eq:dae} and \eqref{eq:theta:ztos} is provided in Section~\ref{deriv:theta} of the Appendix. 
Note that \eqref{eq:theta} in fact describes a family of \ac{tdi} methods.  Methods commonly used in power system simulation software arise from \eqref{eq:theta} as special cases.  For example, $\theta=0.5$ gives the \ac{tm}, while $\theta=0$ gives the \ac{bem}.


Given the value of $(\bfg x_{n}, \bfg y_{n})$
at some time in the simulation, the method computes at each step the new value $(\bfg x_{n+1}, \bfg y_{n+1})$, which is an approximation of the exact solution of \eqref{eq:dae}, i.e.:
\begin{equation}
\begin{aligned}
\label{eq:discrxy}
\bfg x_{n+1-\ell} \approx \bfg x(t+(1-\ell) h) \, , \\
\bfg y_{n+1-\ell} \approx \bfg y(t+(1-\ell) h) \, ,
\end{aligned}    
\end{equation}
where 
$\bfg x_{n+1-\ell} : \mathbb{N} \rightarrow\mathbb{R}^{\nx}$, $\bfg y_{n+1-\ell} : \mathbb{N} \rightarrow \mathbb{R}^{\ny}$
are the discretized state and algebraic variables, respectively, at time $t+(1-l)h$, with
$l \in \mathbb{N}$ and $n=t/h$.
The solution $(\bfg x_{n+1}, \bfg y_{n+1})$ at each step is typically obtained through Newton iterations.

\subsection{DDAE Power System Model}

If time delays are present in the system, the set of \acp{dae} \eqref{eq:dae} changes into a set of \acp{ddae}.  Using a Hessenberg form, we have:
\begin{equation}
  \begin{aligned}
    \label{eq:ddae}
    {\bfg x}' &= \bfg f(\bfg x, \bfg y, 
    \bfg x_d, \bfg y_d)   
    \, , 
    \\
\bfg 0_{\ny,1} &=  \bfg g(\bfg x, \bfg y,  
    \bfg x_d, \bfg y_d)  
 \, ,
  \end{aligned}    
\end{equation}
with:
\begin{equation}
\begin{aligned}
\label{eq:xy_delayed}
    \bfg x_d &= 
    \{ \bfg x(t-\tau_1), \bfg x(t-\tau_2), \ldots, \bfg x(t-\tau_m) \}\, , \\
     \bfg y_d &= 
    \{ \bfg y(t-\tau_1), \bfg y(t-\tau_2), \ldots, \bfg y(t-\tau_m) \}\, ,
\end{aligned}
\end{equation}
where $\tau_i >0$, $i=1,2,\ldots,m$, is the $i$-th time delay of the system and $m$ is the total number of delays.
Note that \eqref{eq:ddae} is not index-1 \cite{book:eigenvalue}. 
%
Under the assumption that $\bfg g_y = \partial\bfg g/\partial\bfg y$ is not singular at every $t$, an index-1 \ac{ddae} model can be constructed if it is additionally assumed that retarded algebraic variables do not appear in the algebraic equations.  In this paper we refrain from this assumption and work with the more general system \eqref{eq:ddae}.
We also note that \eqref{eq:ddae} assumes a constant delay model, which suffices to study the impact on numerical stability and comes without loss of generality.  A detailed discussion on modeling and analysis of systems with time-varying delays that include noise, periodicity and data packet dropouts, can be found in \cite{liu2018stability, tzounas_damping}.
%
%
%


We proceed to provide the formulation of the Theta method for the solution of system \eqref{eq:ddae}.  Applying the Laplace transform to \eqref{eq:ddae} and omitting for simplicity the initial conditions:
\begin{equation}
  \begin{aligned}
    \label{eq:laplace:ddae}
    s \mathcal L\{ {\bfg x} \} 
    &= 
    \mathcal L\{
    \bfg f(\bfg x, \bfg y, \bfg x_d, \bfg y_d) \}
    \, , 
    \\
\bfg 0_{\ny,1} &=  \mathcal L\{
    \bfg g(\bfg x, \bfg y, \bfg x_d, \bfg y_d)\}
     \, .
  \end{aligned}    
\end{equation}
%

We first consider the simplest case where a single 
delay $\tau$: ~$kh \leq \tau < (k+1)h, \; k \in \mathbb{N}$, is present in the system, i.e.:
\begin{equation}
  \begin{aligned}
   \label{eq:xy_d_single}
    \bfg x_d &= 
     \bfg x(t-\tau) , \\
     \bfg y_d &= 
     \bfg y(t-\tau) .
  \end{aligned}
\end{equation}
Delays that are not multiples of the time step, that is $kh < \tau < (k+1)h$, are handled with linear interpolation between $kh$ and $(k+1)h$.  For example, for a given retarded variable $v(t-\tau)$, we use the approximation:
\begin{equation}
  \begin{aligned}
   \label{eq:interpolation}
     v(t-\tau) & \approx  
     c \,
    v(t-kh) + (1-c)\, v(t-(k+1)h) \, ,
    \\
    & \approx  
    c \, v_{n-k}+ (1-c) \, v_{n-(k+1)} \, ,
  \end{aligned} 
\end{equation} 
where the linear interpolation coefficient $c$ is given by:
\begin{align*}
    c = \dfrac{(k+1)h- \tau}{h} \, .
\end{align*}
%
%
%
From \eqref{eq:discrxy}, \eqref{eq:theta:ztos}, we arrive to the following expression:
%
\begin{equation}
  \begin{aligned}
    \label{eq:theta:zdomain:ddae}
    \hspace{-2.2mm}
    {(z - 1)} \mathcal Z\{ {\bfg x_n} \} 
    &= \hspace{-0.3mm}
    h(\theta + (1-\theta) z) \mathcal Z\{
    \bfg f(\bfg x_n, \bfg y_n, 
    \bfg {\mathcal{V}}_{k}^{k+1}
    )\}
    \, , 
    \\
\bfg 0_{\ny,1} 
&= 
\hspace{-0.3mm}
\mathcal Z
\{ \bfg g(\bfg x_{n+1}, \bfg y_{n+1}, 
   \bfg {\mathcal{V}}_{k}^{k+1})
   \} \, ,
  \end{aligned}    
\end{equation}
where $\mathcal{Z}\{ \cdot \}$ denotes the $Z$-transform and
%
\begin{align*} 
\bfg {\mathcal{V}}_{k}^{k+1} &= 
\{ \bfg x_{n-k}, \bfg y_{n-k}, 
\bfg x_{n-(k+1)}, \bfg y_{n-(k+1)} \} \, .
\end{align*}
Applying the inverse $Z$-transform to \eqref{eq:theta:zdomain:ddae}, we obtain:
%
\begin{equation}
  \begin{aligned}
    \label{eq:theta:ddae:single}
   \bfg x_{n+1}  &  = \bfg x_{n} +
    h\theta\bfg f(\bfg x_n, \bfg y_n, 
   \bfg {\mathcal{V}}_{k}^{k+1}
   )
   \\
    &
   \hspace{9mm} + h(1-\theta)
    \bfg f(\bfg x_{n+1}, \bfg y_{n+1}, 
   \bfg {\mathcal{V}}_{k-1}^{k})
    \, , 
    \\
\bfg 0_{\ny,1} &= h
    \bfg g(\bfg x_{n+1}, \bfg y_{n+1}, 
    \bfg {\mathcal{V}}_{k-1}^{k})
   \, .
  \end{aligned}    
\end{equation}
%

Equation \eqref{eq:theta:ddae:single} is the discrete equivalent of \eqref{eq:ddae}, approximated by the Theta method, when only a single delay is present in the system.
This equation can be then conveniently generalized for the multiple-delay case.  
In particular, for a total number of $m$ delays $\tau_i$, with $k_i h \leq \tau_i < (k_i+1)h$, $i=\{ 1,2,\ldots,m\}$ and $k_i \in \mathbb{N}$, application of the 
linear fractional transformation \eqref{eq:theta:ztos} to \eqref{eq:laplace:ddae} yields:
%
\begin{equation}
  \begin{aligned}
    \label{eq:theta:ddae:multiple}
   \bfg x_{n+1}  &  = 
   \bfg x_{n} +
    h\theta\bfg f(\bfg x_n, \bfg y_n, 
   \bfg {\mathcal{V}}_{k_1}^{k_1+1},
   \ldots,
   \bfg {\mathcal{V}}_{k_m}^{k_m+1}
   ) \\
    &
   \hspace{9mm} + h(1-\theta)
    \bfg f(\bfg x_{n+1}, \bfg y_{n+1}, 
   \bfg {\mathcal{V}}_{k_1-1}^{k_1},
   \ldots,
  \bfg {\mathcal{V}}_{k_{m-1}}^{k_m})
    \, , 
    \\
\bfg 0_{\ny,1} &= h
    \bfg g(\bfg x_{n+1}, \bfg y_{n+1}, 
   \bfg {\mathcal{V}}_{k_1-1}^{k_1},
   \ldots,
   \bfg {\mathcal{V}}_{k_{m-1}}^{k_m})
   \, .
  \end{aligned}    
\end{equation}
 
The main concept discussed in this paper is that, being implicit notwithstanding, 
\eqref{eq:theta:ddae:multiple} can perform poorly and even become numerically unstable when applied to a power system model with inclusion of multiple delayed variables.  This is illustrated theoretically through numerical stability analysis, in Section~\ref{sec:num_stab}, as well as through numerical simulations in the case study of Section~\ref{sec:case}.


%
%
%
%
%

\section{Numerical Stability Analysis}
\label{sec:num_stab}

In this section, we first illustrate through numerical stability analysis of a scalar test equation that standard implicit \ac{ode} methods can be destabilized when applied to a time-delayed system.  We then proceed to describe a matrix pencil approach
to study the conditions under which such numerical instabilities can arise in the simulation of realistic power system models. 

\subsection{Analytical Study of Test Delay Differential Equation}
 
The standard approach to study the stability properties of a numerical \ac{tdi} method for
\acp{ode} is to evaluate its response when applied to the linear test differential equation:
\begin{equation}\label{eq:test}
x'(t)=a x(t) \, , 
\ \ a \in \mathbb{C} \, .
\end{equation}
%
Applying the Theta method to \eqref{eq:test} we have:
\begin{equation}\label{eq:stabfunc}
  x_{n+1}=
  \dfrac{{1+h \theta a}}{1-h (1-\theta) a}
  x_{n} \, ,
\end{equation}
where 
\begin{equation}\label{eq:theta:stability}
  \mathcal{R}(a h) =
  \dfrac{{1+h \theta a}}{1-h (1-\theta) a} \, ,
\end{equation}
is the method's \textit{growth function}.  The stability region of the method is defined by the set $\{ a \in \mathbb{C} \, : \ |\mathcal{R}(a h)|<1 \}$.
For example, $\theta=0.5$ gives the \ac{tm}, for which:
\begin{equation}\label{eq:tm:stability}
\mathcal{R}(a h) =
\dfrac{{1 + 0.5 h a}}{1 - 0.5 h a} \, .
\end{equation}
The stability region in this case is ${\rm Re}\{a h\}<0:\ a h \in \mathbb{C}$, i.e.,~the open left half of the \textit{s}-plane (symmetrical A-stability).



Let's consider that a time delay $\tau=h$ is introduced to \eqref{eq:test}.  Then, the latter changes into the following test delay differential equation: 
%
%
\begin{equation}
\label{eq:test:dde}
x'(t) = a x(t) + b x(t-h)  \, ,
\end{equation}
where $a,\,b$ are the coefficients of the delay-free and delayed terms 
of the equation, respectively.
%
%
Application of the Theta method to \eqref{eq:test:dde} gives:
\begin{equation}
 \label{eq:test:dde:growth_function}
 x_{n+1} 
 \hspace{-0.5mm} 
 = 
 \hspace{-0.5mm}
 \dfrac{1 + a h \theta + b h (1-\theta)}{1-a h(1-\theta)}
 x_n 
 + \dfrac{b h \theta}{1-a h(1-\theta)}
 x_{n-1} \, .
\end{equation}
%
%
By setting $\bfb{y}_{n} = [ x_{n},  x_{n-1}]\T$, \eqref{eq:test:dde:growth_function} can be expressed in the form:
\begin{equation}
 \label{eq:test:dde:growth_matrix}
  \bfb{y}_{n+1} = \bfb{\mathcal{R}} \; \bfb{y}_{n} \; ,
\end{equation}
where:
\begin{equation}
 \label{eq:growth_matrix}
  \bfb{\mathcal{R}} =
    \begin{bmatrix}
      \dfrac{1 + a h \theta + b h (1-\theta)}{1-a h(1-\theta)} & \dfrac{b h \theta}{1-a h(1-\theta)} \vspace{2mm} \\
      1 & 0
    \end{bmatrix} .
\end{equation}
The stability region of the Theta method when applied to \eqref{eq:test:dde} can thus be determined through the eigenvalues of matrix $\bfb{\mathcal{R}}$.  In particular, the method is stable in the region for which $\rho(\bfb{\mathcal{R}})<1$, where $\rho(\cdot)$ denotes the spectral radius of a matrix.  

\begin{figure}[ht!]
  \centering
  \begin{subfigure}{0.49\columnwidth}
    \centering
    \includegraphics[width=\linewidth]{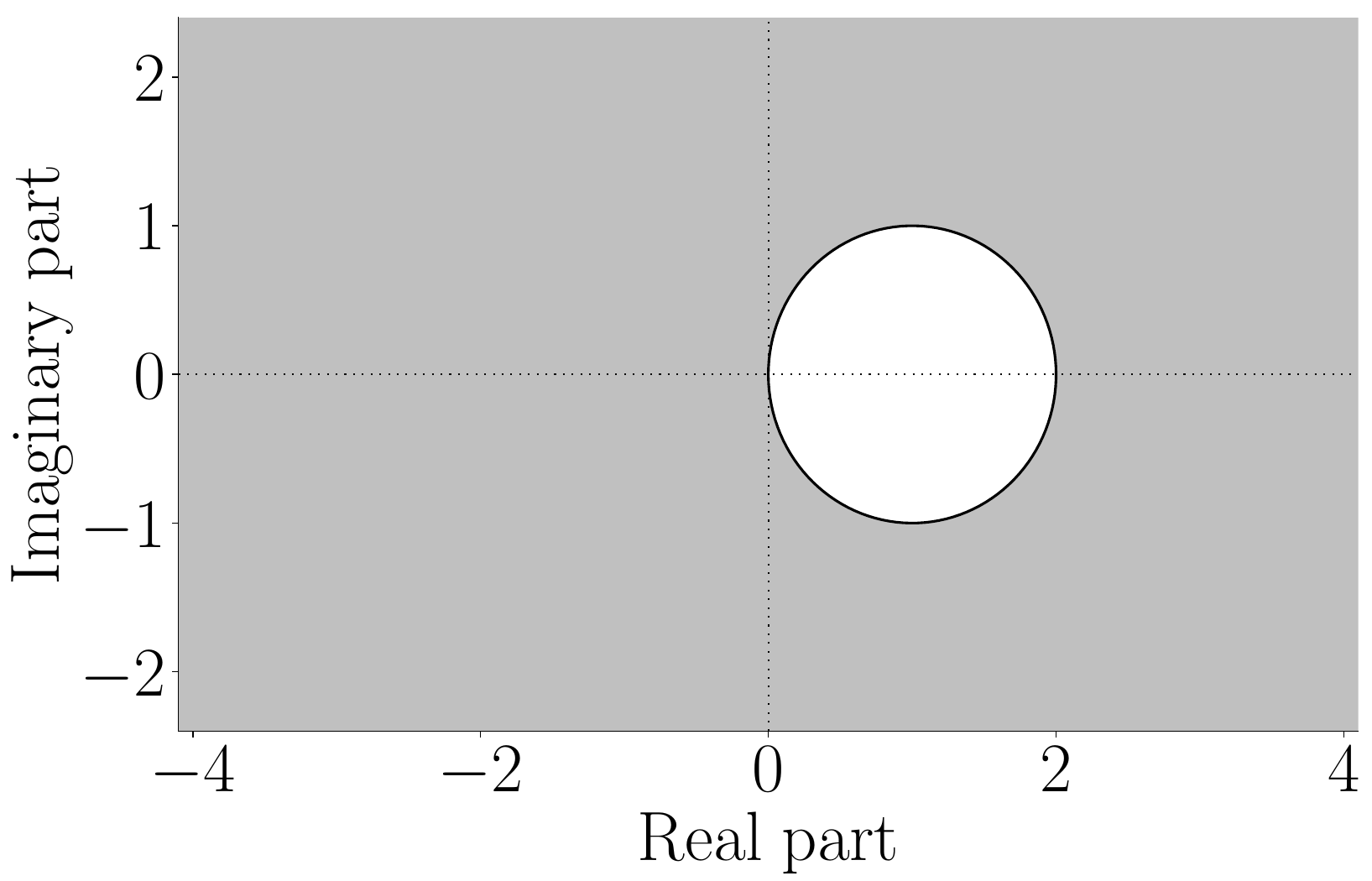}
    \caption{$x'=a \; x(t)$}
    \label{fig:bem1}
  \end{subfigure}
  \hfill
  \begin{subfigure}{0.49\columnwidth}
    \centering
    \includegraphics[width=\linewidth]{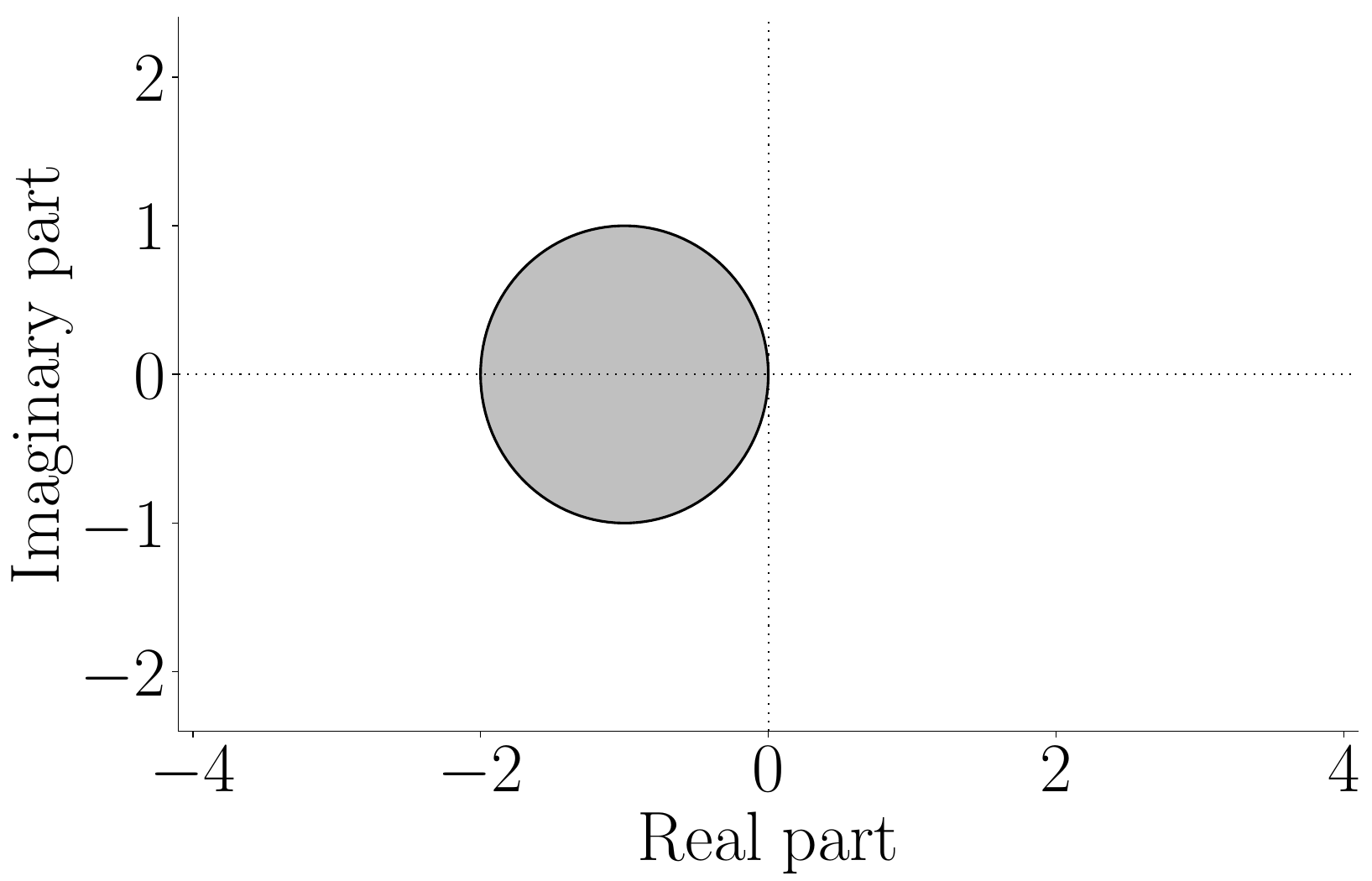}
    \caption{$x'=b \; x(t-h)$}
    \label{fig:bem2}
  \end{subfigure}
  \begin{subfigure}{0.49\columnwidth}
  \vspace{3mm}
    \centering
    \includegraphics[width=\linewidth]{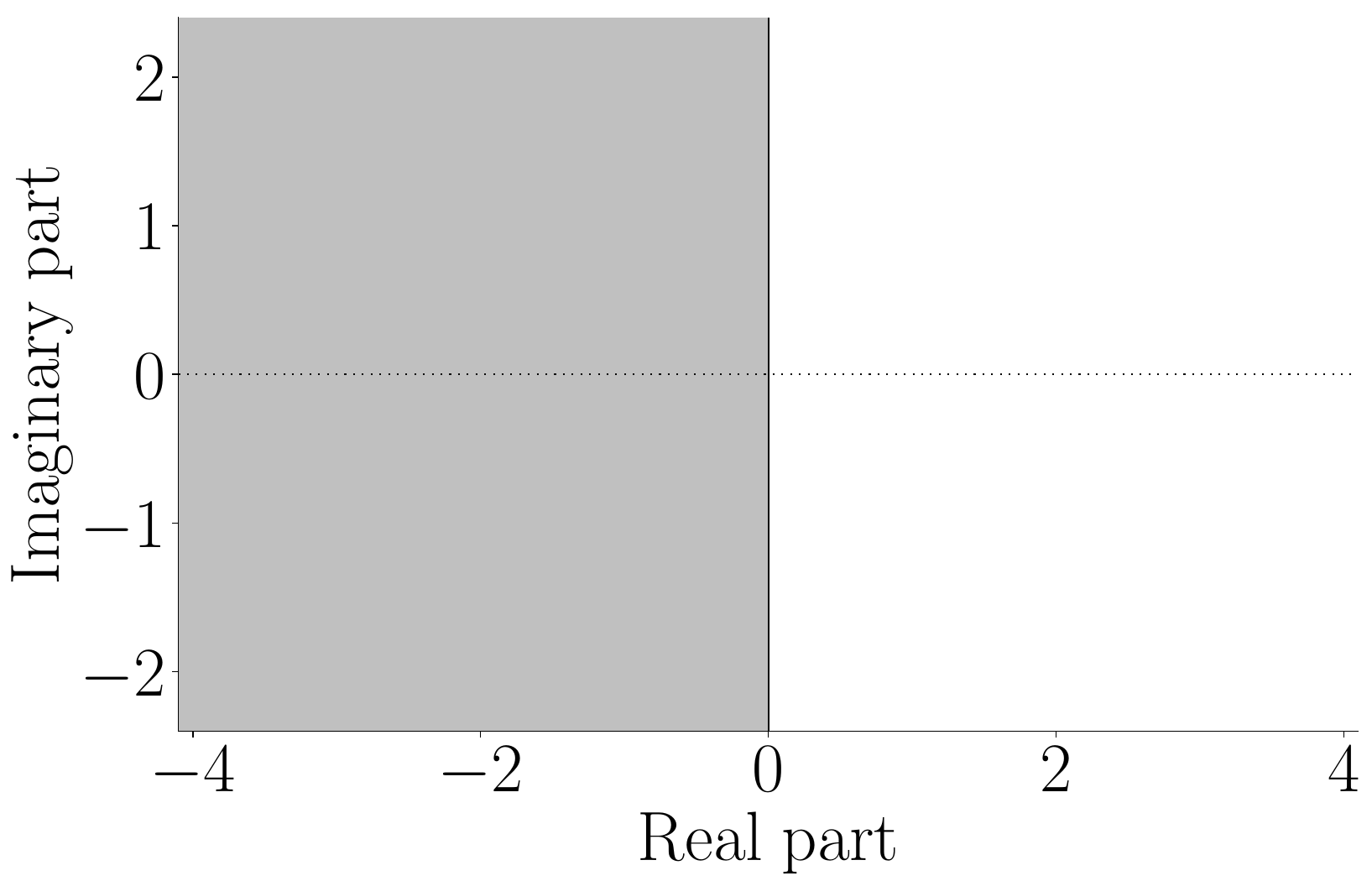}
    \caption{$x'=a \;  x(t) + a \;  x(t-h)$}
    \label{fig:bem3}
  \end{subfigure}
  \hfill
  \begin{subfigure}{0.49\columnwidth}
    \centering
    \includegraphics[width=\linewidth]{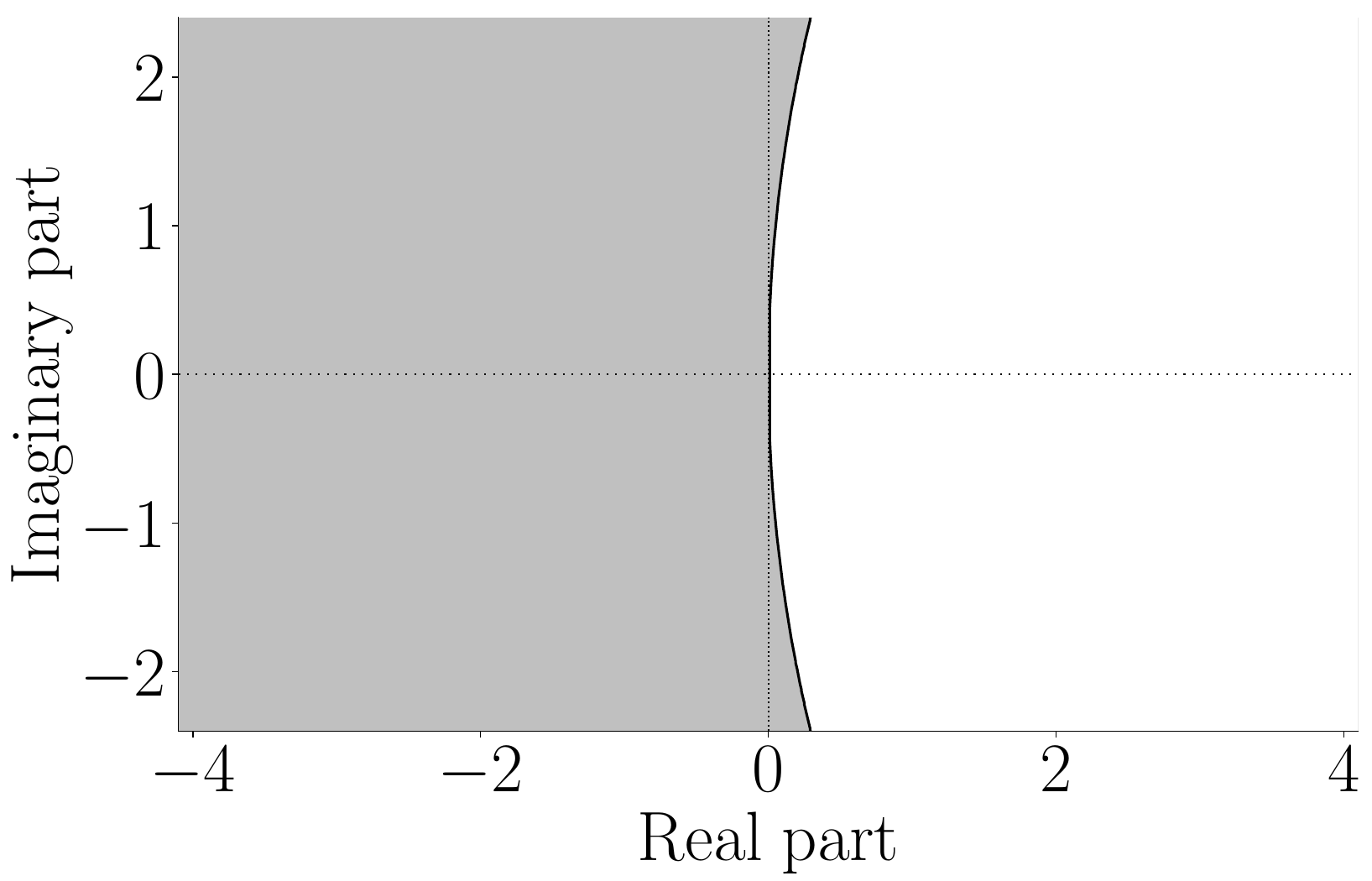}
    \caption{$x'=1.1b \; x(t) + b \; x(t-h)$}
    \label{fig:bem4}
  \end{subfigure}
   \begin{subfigure}{0.49\columnwidth}
   \vspace{3mm}
    \centering
    \includegraphics[width=\linewidth]{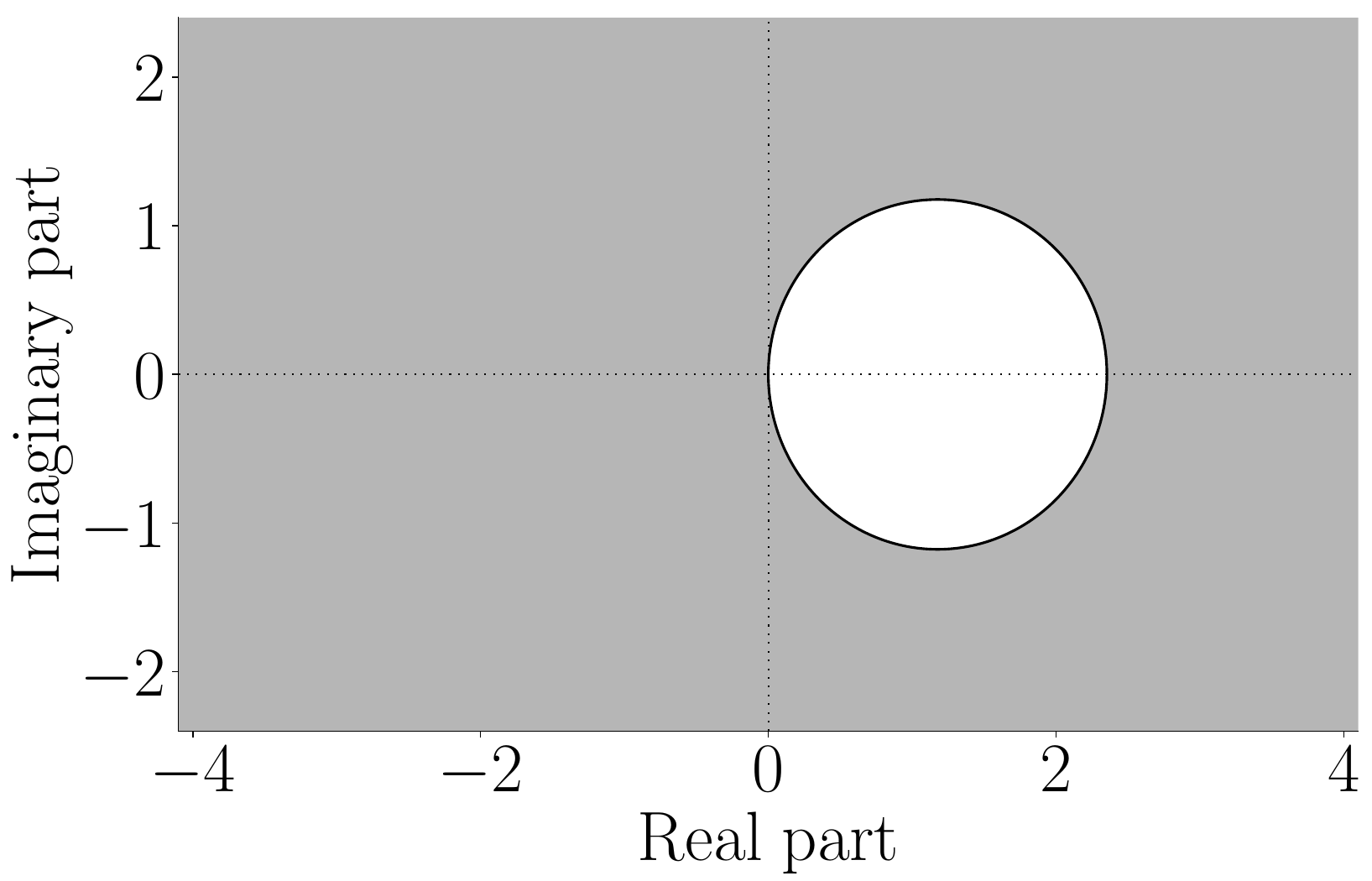}
    \caption{$x'=a \; x(t) + 0.15a \; x(t-h)$}
    \label{fig:bem5}
  \end{subfigure}
  \hfill
  \begin{subfigure}{0.49\columnwidth}
    \centering
    \includegraphics[width=\linewidth]{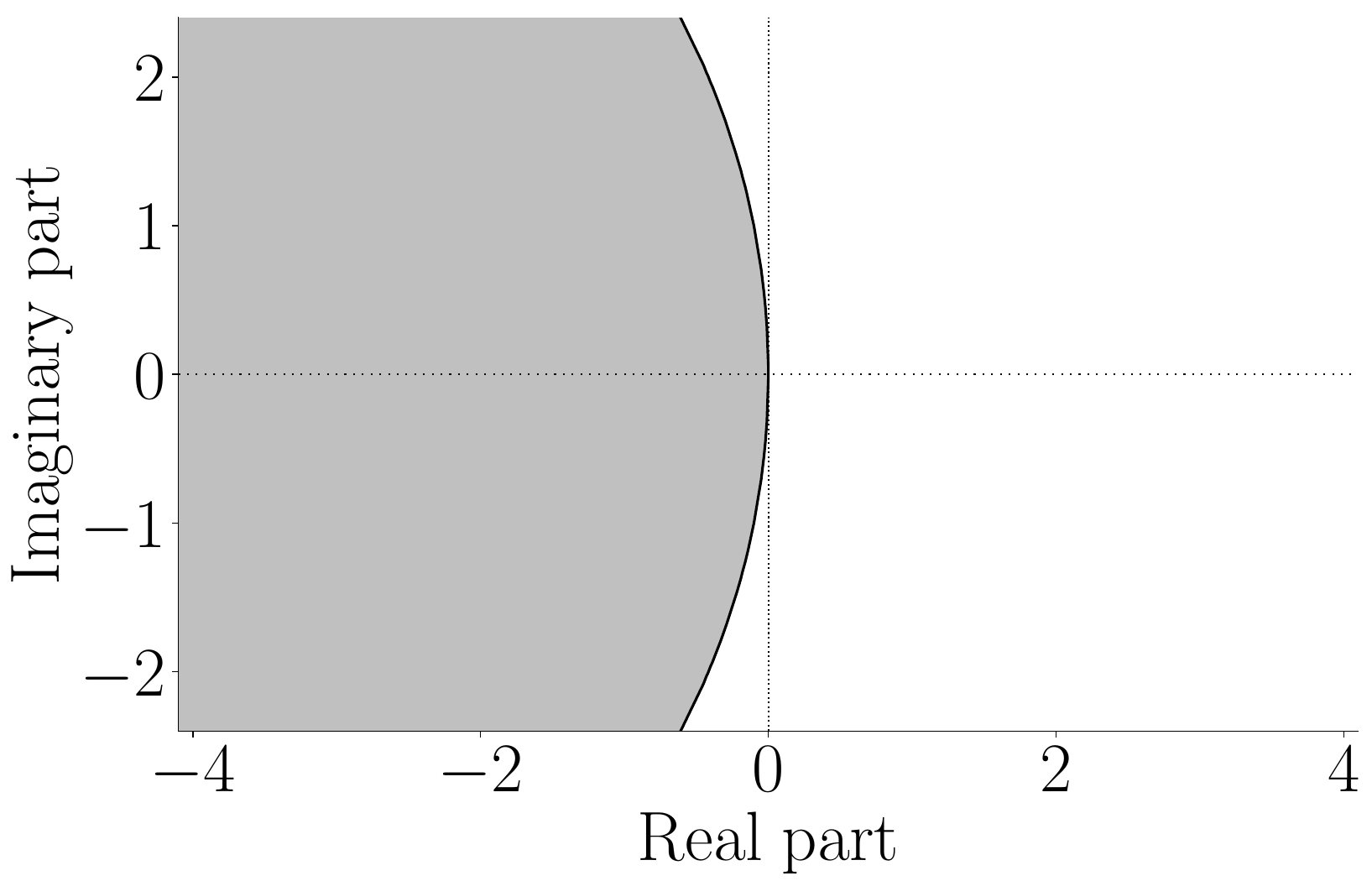}
    \caption{$x'=0.85b \; x(t) + b \; x(t-h)$}
    \label{fig:bem6}
  \end{subfigure}
  \caption{Stability region for test equation \eqref{eq:test:dde}, $\theta = 0$ (\ac{bem}).}
  \label{fig:bem_scalar}
\end{figure}
\begin{figure}[ht!]
  \centering
  \begin{subfigure}{0.49\columnwidth}
    \centering
    \includegraphics[width=\linewidth]{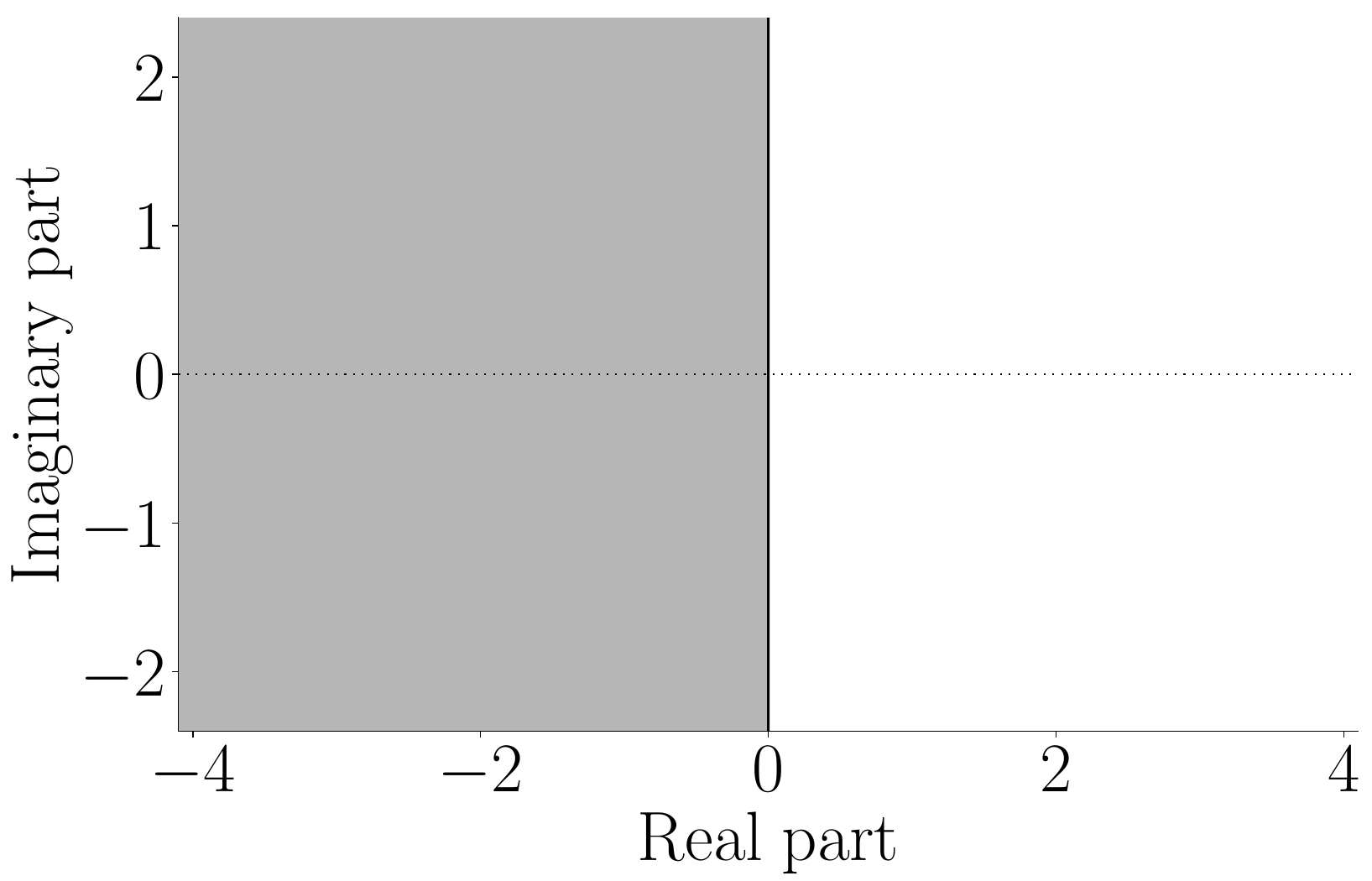}
    \caption{$x'=a \; x(t)$}
    \label{fig:tm1}
  \end{subfigure}
  \hfill
  \begin{subfigure}{0.49\columnwidth}
    \centering
    \includegraphics[width=\linewidth]{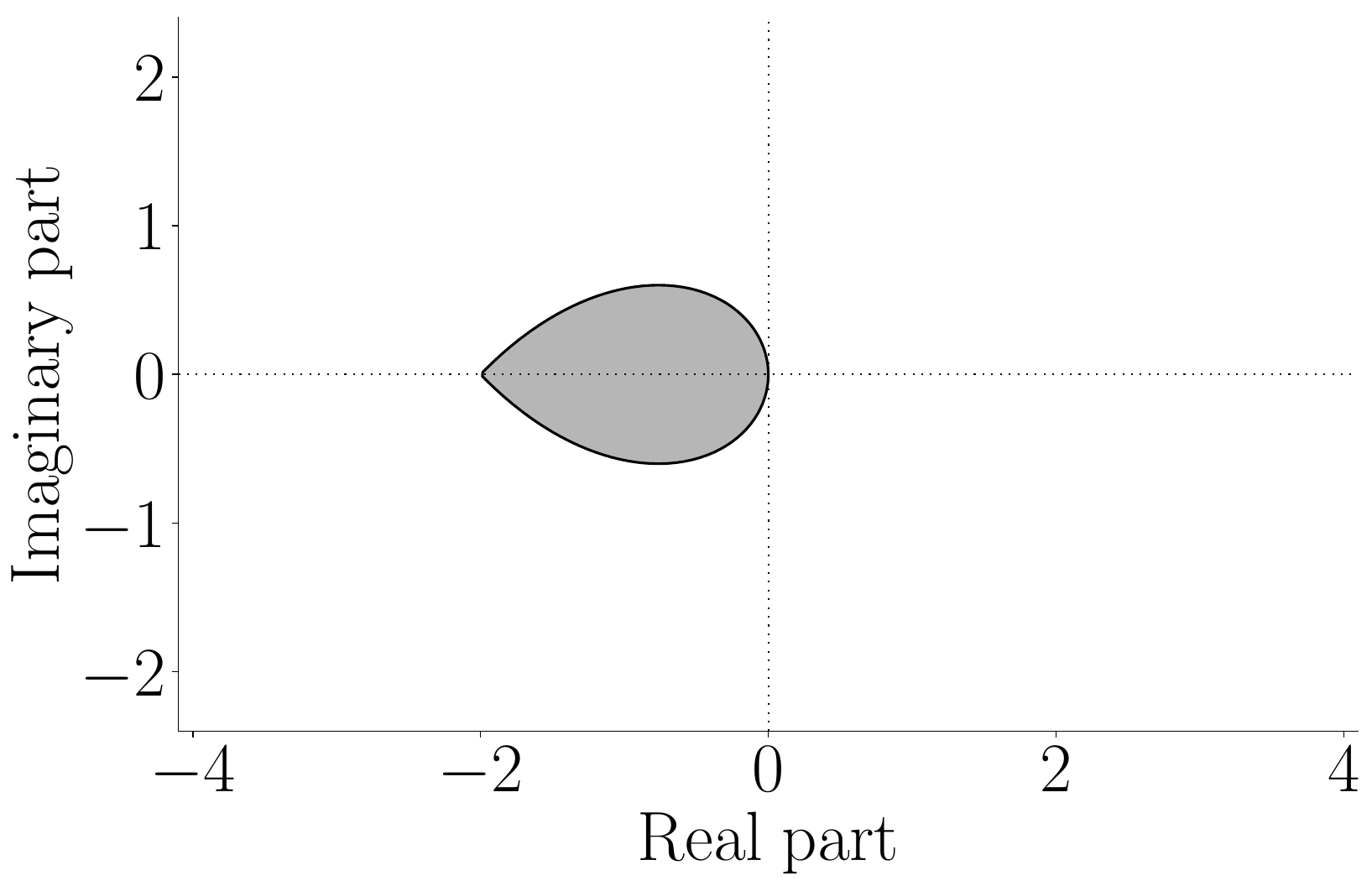}
    \caption{$x'=b \;  x(t-h)$}
    \label{fig:tm2}
  \end{subfigure}
  \begin{subfigure}{0.49\columnwidth}
  \vspace{3mm}
    \centering
    \includegraphics[width=\linewidth]{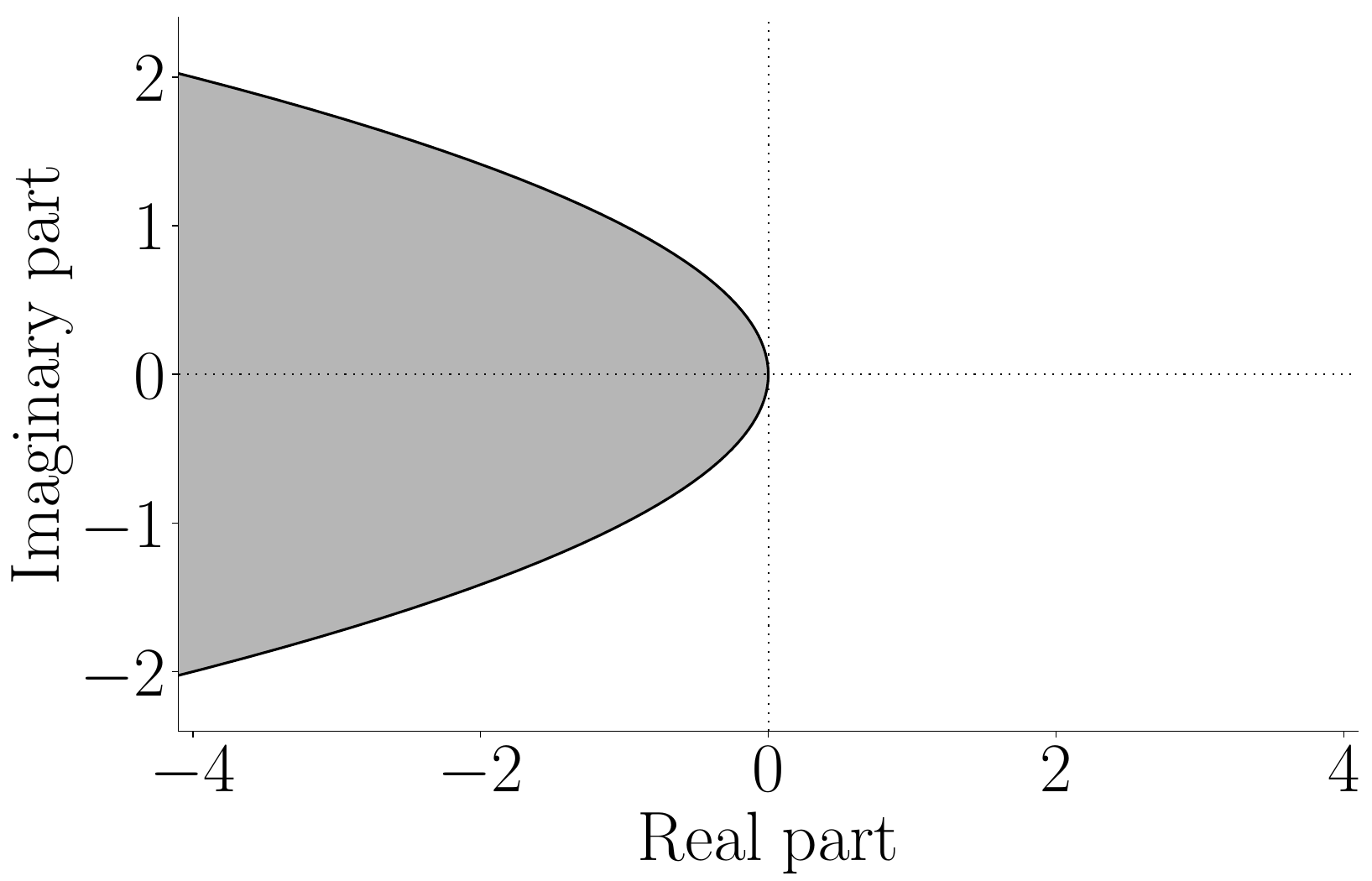}
    \caption{$x'=a  \; x(t) + a \; x(t-h)$}
    \label{fig:tm3}
  \end{subfigure}
  \hfill
  \begin{subfigure}{0.49\columnwidth}
    \centering
    \includegraphics[width=\linewidth]{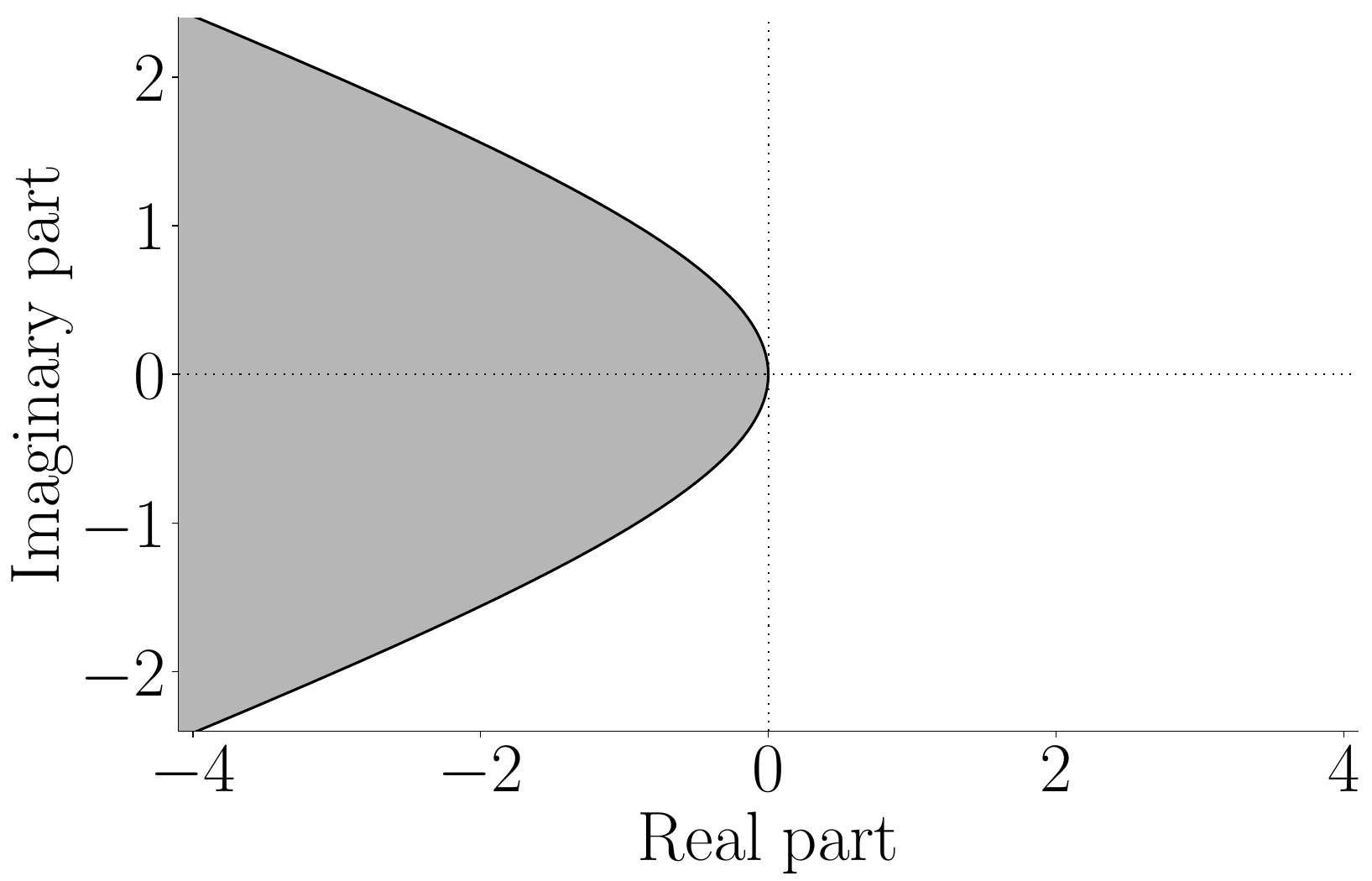}
    \caption{$x'=1.1b \; x(t) + b \; x(t-h)$}
  \label{fig:tm4}
  \end{subfigure}
   \begin{subfigure}{0.49\columnwidth}
   \vspace{3mm}
    \centering
    \includegraphics[width=\linewidth]{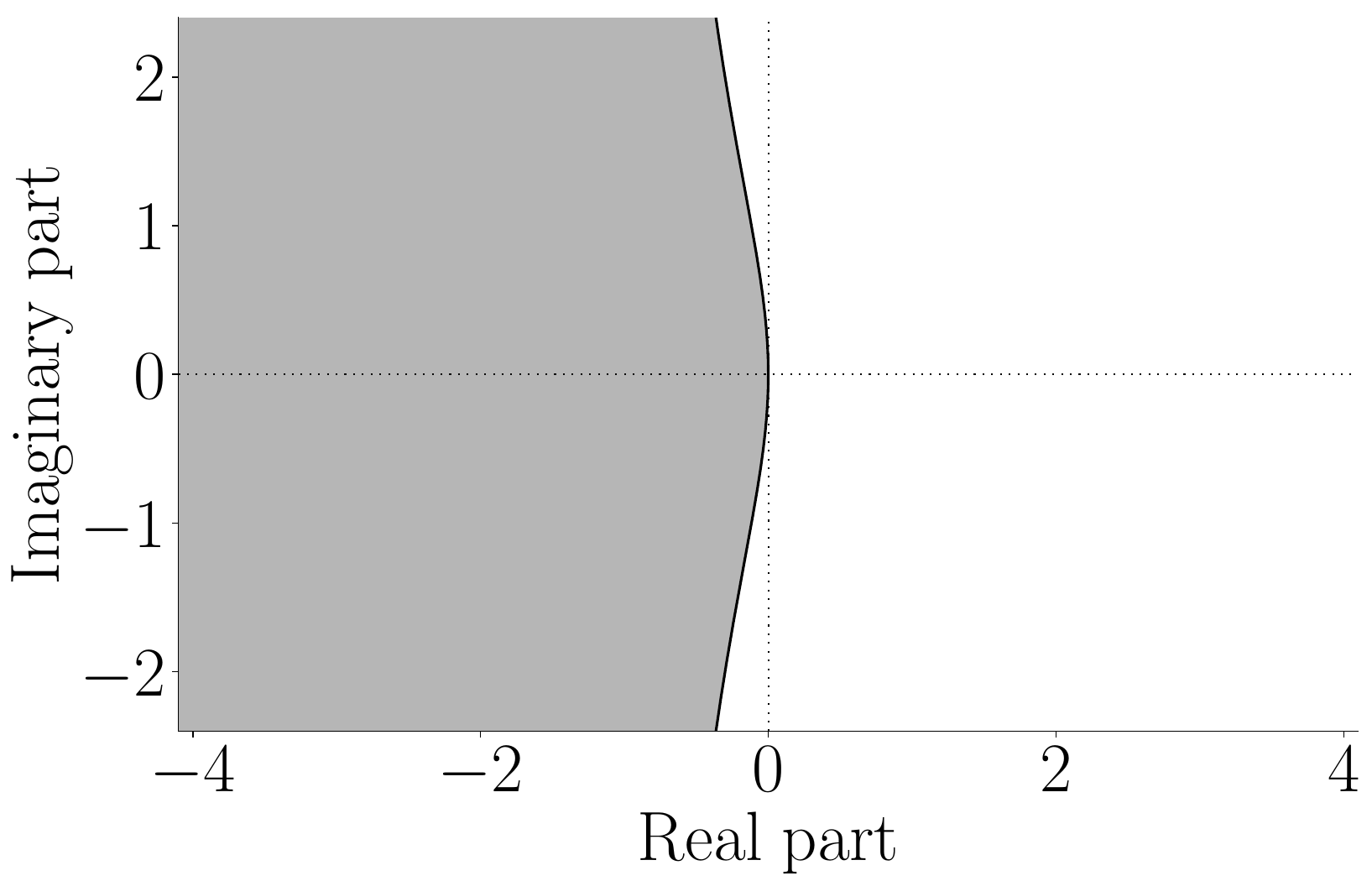}
\caption{$x'=a \; x(t) + 0.15a \; x(t-h)$}
\label{fig:tm5}
\end{subfigure}
\hfill
\begin{subfigure}{0.49\columnwidth}
\centering
\includegraphics[width=\linewidth]{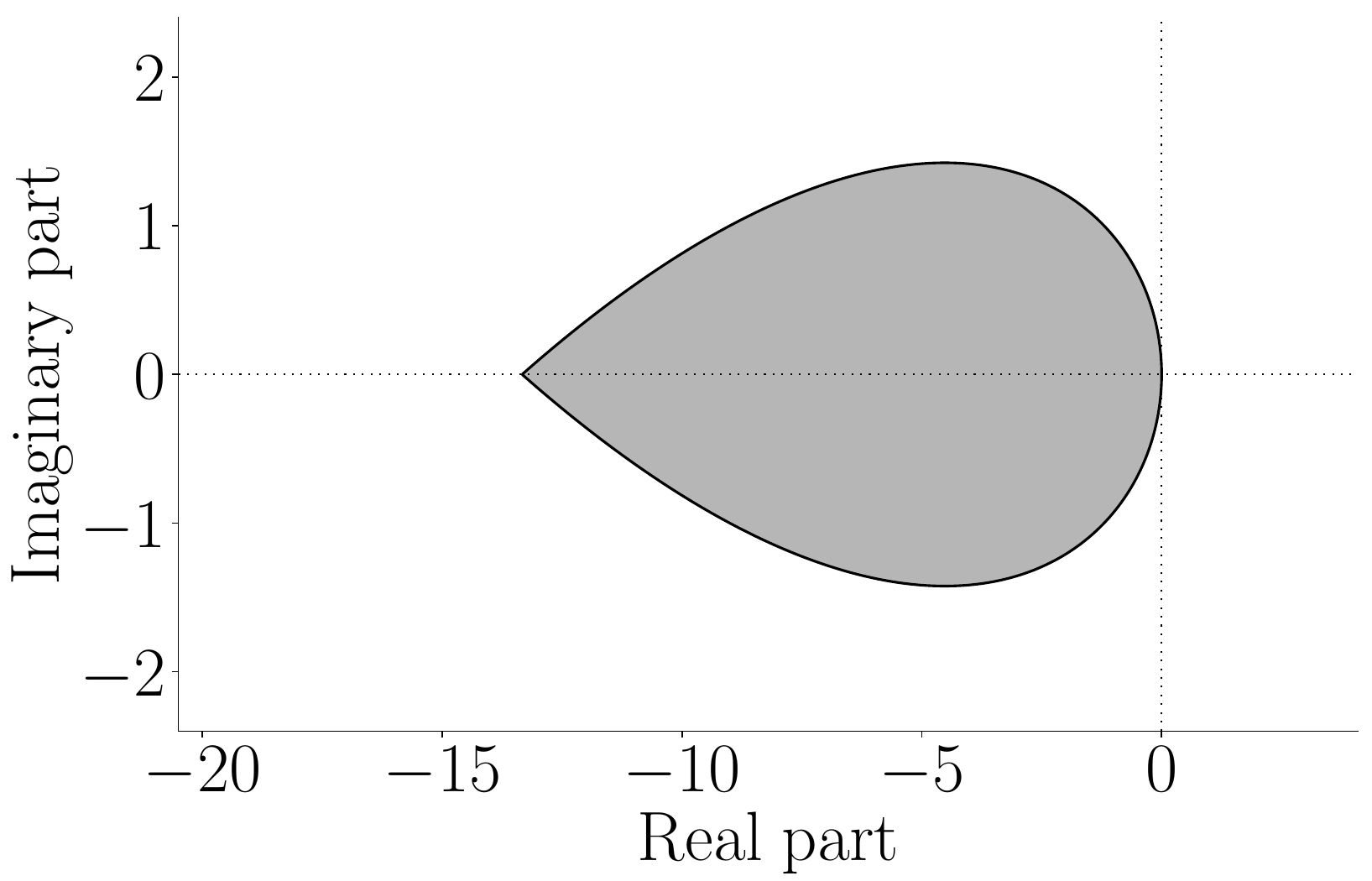}
  \caption{$x'=0.85bx(t) + bx(t-h)$}
  \label{fig:tm6}
  \end{subfigure}
  \caption{Stability region for test equation \eqref{eq:test:dde}, $\theta = 0.5$ (\ac{tm}).}
  \label{fig:tm_scalar}
\end{figure}

As already discussed in Section~\ref{sec:model}, 
two important elements of the family of integration schemes that the Theta method covers are obtained for $\theta=0$ (\ac{bem}) and $\theta=0.5$ (\ac{tm}). Figures~\ref{fig:bem_scalar} and \ref{fig:tm_scalar} illustrate the stability regions of these methods when applied to \eqref{eq:test:dde}, considering different weights $a$ and $b$ for the delay-free and delayed parts of the equation, respectively.
In particular, 
Figs.~\ref{fig:bem1} and \ref{fig:tm1} show the stability regions for $b=0$, in which case \eqref{eq:test:dde} is reduced to the test \ac{ode} \eqref{eq:test}.  As it is well-known, the stability region of the \ac{bem} for \acp{ode} lies outside the unit circle centered at $(1,0)$ of the complex plane, and the method is L-stable.  Moreover, the stability region of the \ac{tm} for \acp{ode} is the open left half of the complex plane, and the method is symmetrically A-stable. 

The above important 
stability properties do not extend to time-delayed systems.  The extreme theoretical scenario is $a=0$, in which case the evolution of the present state in \eqref{eq:test:dde} is influenced solely by the past value $x(t-h)$.  In this scenario, both \ac{bem} and \ac{tm} are \textit{unstable} methods.  As a matter of fact, the stability region of the \ac{bem} in this case is identical to that of the explicit Euler method for \acp{ode}.  Then, the plots in Figs.~\ref{fig:bem2}-\ref{fig:bem6} basically illustrate that the \ac{bem} is numerically stable only if $a\geq b$.  Most importantly, Figs.~\ref{fig:tm2}-\ref{fig:tm6} indicate that, for $b \neq 0$, there exists a time step size $h>0$ above which the \ac{tm} is destabilized.  In fact, the stability region of the method shrinks as the ratio $b/a$ is increased, i.e.,~the risk for numerical instability becomes higher, as the influence of the past state on $x'(t)$ gets stronger. 

Extrapolating these observations to a time-delayed system, one can expect that the number of delayed variables and the magnitudes of their coefficients have a crucial impact on the stability region of the Theta method applied.  Then, in power system models with a large number of non-negligible delayed variables, there are expected to be cases for which e.g., the \ac{tm} significantly deforms the system's dynamic modes even for reasonably small time step sizes, potentially making stable trajectories appear as unstable.  Needless to say, the specific conditions under which such numerical issues arise will be system dependent. The question is then how to study the numerical deformation introduced by the integration method to the dynamics of a given power system model that includes multiple delayed variables. This is discussed in detail in the next paragraph.

\subsection{Systems with Multiple Time-Delayed Variables}

For a scalar test equation, such as \eqref{eq:test:dde} discussed above, one can characterize analytically the properties of Theta method by identifying its numerical stability region (e.g.,~see Figs.~\ref{fig:bem_scalar}, \ref{fig:tm_scalar}).  However, real-world power system models are too large in size and too complex to permit calculation of numerical \ac{tdi} stability regions.  
In this section, we address this limitation of standard numerical stability analysis techniques, by describing a \ac{sssa}-based approach to capture the impact of delays on the behavior of the Theta method when applied to a dynamic power system model.


Consider the \ac{ddae} power system model \eqref{eq:ddae} and assume that a stationary solution $(\bfg x_o,\bfg y_o)$ is known.  
Then, differentiating \eqref{eq:ddae} at the stationary point yields:
%
\begin{equation}
\begin{aligned}
\label{eq:lin:ddae}
\tilde{\bfg x}' 
&= \jac{f}{x} \tilde{\bfg x}  + \jac{f}{y} \tilde{\bfg y}
+ \sum_{k=1}^\nd [\jac{f}{x,k}  \tilde{\bfg x}(t- k h) 
+ \jac{f}{y,k}  \tilde{\bfg y}(t-k h) ] \, ,\\
\bfg 0_{\ny,1} 
& = \jac{g}{x} \tilde{\bfg x}  +
\jac{g}{y} \tilde{\bfg y}
+ \sum_{k=1}^\nd [ \jac{g}{x,k} \tilde{\bfg x}(t-k h) + \jac{g}{y,k} \tilde{\bfg y}(t-k h) ] \, .
\end{aligned}
\end{equation}
%
In \eqref{eq:lin:ddae}, delays appear as multiples of the time step as the delay-free ($\jac{f}{x}$, $\jac{f}{y}$, $\jac{g}{x}$, $\jac{g}{y}$) and delayed ($\jac{f}{x,k}$, $\jac{f}{y,k}$, $\jac{g}{x,k}$, $\jac{g}{y,k}$)
Jacobian matrices are formulated to account for delay interpolation according to \eqref{eq:interpolation}.  Moreover, the upper limit of summation $r$ is 
such that  $ r \in \mathbb{N^+}$: $(\nd-1)h<\tau_{\max} \leq \nd h$, where $\tau_{\max}$ is the delay with the largest magnitude.

By setting $\tilde{\bfg \xs} = (\tilde{\bfg x}, \tilde{\bfg y})$, \eqref{eq:lin:ddae} becomes equivalently:
\begin{equation}
\begin{aligned}
\label{eq:lin_matrix:ddae}
\bfb E \tilde{\bfg \xs}' &= \bfb A_{0} \tilde{\bfg \xs} + \sum_{k=1}^\nd \bfb A_{k}  \tilde{\bfg \xs}(t- k h) \, ,
\end{aligned}
\end{equation}
where:
\begin{align*}
\bfb E =
    \begin{bmatrix}
\bfg I_{\nx} & \bfg 0_{\nx,\ny} \\
\bfg 0_{\ny,\nx} & \bfg 0_{\ny,\ny}
    \end{bmatrix} 
    , \, 
    \bfb A_{0} = 
     \begin{bmatrix}
      \jac{f}{x}  & \jac{f}{y} \\
\jac{g}{x} & \jac{g}{y}
    \end{bmatrix} 
 , \,
  \bfb A_{k} =    
   \begin{bmatrix}
      \jac{f}{x,k}  & \jac{f}{y,k} \\
\jac{g}{x,k} & \jac{g}{y,k} 
    \end{bmatrix} ,
\end{align*}
where $\bfg I_{\nx}$ is the identity matrix of size $\nx$. 
The eigenvalues of \eqref{eq:lin_matrix:ddae} are the roots of the determinant of the system's matrix pencil, which is defined as follows \cite{book:eigenvalue}: 
\begin{equation}
\label{eq:ddae_pencil}
    {s}\bfb E - \bfb A_{0} - \sum_{k=1}^\nd \bfb A_{k} e^{-{s} h k} 
\end{equation}
We note that the time delays present in \eqref{eq:lin:ddae} give rise to exponential terms in \eqref{eq:ddae_pencil}. This suggests the existence of infinitely many characteristic roots \cite{book:eigenvalue}, the calculation of which is prohibitive.  There are various approaches in the existing literature to overcome this issue. The experience of the authors is that arguably the most effective one is to compute an approximately equivalent linear pencil through the application of spectral discretization. 
The procedure includes two steps: first, the linearized DDAE system is transformed into an equivalent system of partial differential equations~(PDEs); then, the PDE system is reduced to a linear eigenvalue problem of finite dimensions, through Chebyshev polynomials.
The detailed description of the spectral discretization technique followed in this paper can be found in \cite{Li_discr}.

Applying the Theta method to \eqref{eq:lin:ddae}:
\begin{align*}
\tilde{\bfg x}_{n+1} &= 
\tilde{\bfg x}_{n} +
h(1-\theta) 
(
\jac{f}{x} \tilde{\bfg x}_{n+1}  + 
\jac{f}{y} \tilde{\bfg y}_{n+1}
)
\\
& 
\hspace{7mm}
+ h \theta 
(
\jac{f}{x} \tilde{\bfg x}_{n}  + 
\jac{f}{y} \tilde{\bfg y}_{n}
) \\
& 
\hspace{7mm} + 
h(1-\theta) 
\sum_{k=1}^\nd ( \jac{f}{x,k}  
\tilde{\bfg x}_{n+1-k} 
+ \jac{f}{y,k} \tilde{\bfg y}_{n+1-k} )
\\
&
\hspace{7mm} + 
h\theta 
\sum_{k=1}^\nd ( \jac{f}{x,k}  
\tilde{\bfg x}_{n-k} 
+ \jac{f}{y,k} \tilde{\bfg y}_{n-k} ) \, ,
\\
\bfg 0_{\ny,1} & =  h[\jac{g}{x} \tilde{\bfg x}_{n+1}  +
\jac{g}{y} \tilde{\bfg y}_{n+1}]
\\ & 
\hspace{7mm}
+ h \sum_{k=1}^\nd ( \bfg g_{x,k} \tilde{\bfg x}_{n+1-k} 
+ \bfg g_{y,k} \tilde{\bfg y}_{n+1-k} ) \, ,
\end{align*}
or, equivalently:
\begin{align}
\label{eq:theta:ABC}
   \bfb M  \xs_{n+1}
    &= \bfb A  \xs_{n}
    + \sum_{k=1}^\nd
    \left (
    \bfb B_{k-1}  \xs_{n-(k-1)}
    + \bfb C_{k}  \xs_{n-k}
    \right ) \, ,
\end{align}
where $\xs_n = (\tilde{\bfg x}_{n}, \tilde{\bfg y}_{n})$ and: 
\begin{align*}
\bfb M &=
\begin{bmatrix}
\bfg I_{\nx} 
- h (1-\theta)\jac{f}{x} & -h(1-\theta) 
\jac{f}{y} \\
-h\jac{g}{x} & -h\jac{g}{y}
    \end{bmatrix} ,  \\
    \bfb A &= 
     \begin{bmatrix}
\bfg I_{\nx} 
 + h \theta \bfg f_x  & h \theta 
\jac{f}{y} \\
\bfg 0_{\ny,\nx}  
&
\bfg 0_{\ny,\ny} 
    \end{bmatrix} , \\
  \bfb B_{k-1} &=    
   \begin{bmatrix}
h(1-\theta)\jac{f}{x,k} & 
h(1-\theta)\jac{f}{y,k}  \\
h \jac{g}{x,k}  & h \jac{g}{y,k} 
    \end{bmatrix} , \\
 \bfb C_{k} &=    
   \begin{bmatrix}
h\theta \jac{f}{x,k} & 
h\theta \jac{f}{y,k}  \\
\bfg 0_{\ny,\nx} 
&
\bfg 0_{\ny,\ny} 
    \end{bmatrix} \, ,
\end{align*}
which can be rewritten in the form:
\begin{align}
\label{eq:theta:FG}
    \bfb F \, \bfb y_{n+1} &= \bfb G \, \bfb y_n \, .
\end{align}
The proof of \eqref{eq:theta:FG} is provided in Section~\ref{proof:FG} of 
the Appendix.

Equation \eqref{eq:theta:FG} is a discrete-time approximation of \eqref{eq:lin_matrix:ddae}, where $\Etdi$ and $\Atdi$ vary for different values of $\theta$ but are always matrix functions of $\AS$ and $h$. The matrix pencil of \eqref{eq:theta:FG} is:
%
\begin{align}
\label{eq:theta:FG_pencil}
    \hat{z} \bfb F - \bfb G
    \, .
\end{align}
We quantify the numerical deformation introduced by the Theta method to the power system dynamics by comparing the eigenvalues of \eqref{eq:theta:FG_pencil} with those of 
\eqref{eq:ddae_pencil}.  
In this regard, note that the eigenvalues of \eqref{eq:theta:FG_pencil} lie in the \textit{z}-domain whereas those of 
\eqref{eq:lin:ddae} in the \textit{s}-domain. Thus, to make the results comparable, we transform the roots of \eqref{eq:theta:FG_pencil} to the \textit{s}-domain, through the logarithmic transformation:
\begin{align}
\label{eq:log_trans}
    \hat{s}\,=\,\frac{1}{h} \ln{\hat{z}} \, ,
\end{align}
where $\hat{s}$ and $\hat{z}$ represent the deformed eigenvalues in the \textit{s}-domain and \textit{z}-domain respectively.  
Numerical instability is identified when a deformed eigenvalue $\hat{s}$ lies in the positive half of the complex plane, while the real parts of all finite eigenvalues of \eqref{eq:lin:ddae} remain negative.
The approach above is based on \ac{sssa} and hence the results are in principle valid only in a neighborhood around equilibria.   
Nevertheless, when numerical instability is predicted, it is guaranteed to occur, as instability observed for small disturbances inherently implies instability under larger disturbances as well.  
Moreover, the structure and stiffness of power systems as well as the properties of \ac{tdi} methods are features that tend to be robust with respect to the operating condition, and thus the 
results provide a tentative yet accurate estimate of numerical deformation also for varying operating conditions. Similar considerations can be found in the literature, e.g.,~see \cite{tzounas2022tdistab, book:chow:13, arriaga:82_2}. 

\section{Case Study}
\label{sec:case}

This section presents simulation results based on the IEEE 39-bus system, detailed data of which can be found in \cite{web:39bus}.
The original system consists of 10 \acp{sm}, modeled by fourth order, two-axis models, 34 transmission lines, 12 transformers and 19 loads.  Each machine is equipped with an \ac{avr}, a \ac{tg} and a \ac{pss}. Moreover, \acp{sm} are assumed to provide secondary frequency regulation through an \ac{agc} system modeled as an integral controller.  The resulting model has in total 393 variables (131 states and 262 algebraic variables).    
Simulation results in this section are produced using the the Python-based power system analysis software tool Dome \cite{vancouver}.

\subsection{Delay-Free System}
\label{sec:case:delayfree}

We first consider the system without delays.
The stiffness of the examined system is measured through the ratio 
$\mathcal{S} = |s^{\max}|/|s^{\min}|$ between the largest and smallest eigenvalue magnitudes $|s^{\max}|$ and $|s^{\min}|$ of the corresponding linearized model.  
In our case, $\mathcal{S} = 5300.53$. 
The examined system is small-signal stable and the two rightmost pairs of eigenvalues are $s_{1,2}=-0.694176\pm j0.808851$ with damping ratio $\zeta_{1,2}=65.1\%$ and
${s}_{3,4} = -0.013359 \pm j0.050441$ with 
damping ratio $\zeta_{3,4}=25.6\%$.

\begin{figure}[ht!]
  \centering
  \begin{subfigure}{1.0\columnwidth}
    \centering    
    \includegraphics[width=\linewidth]{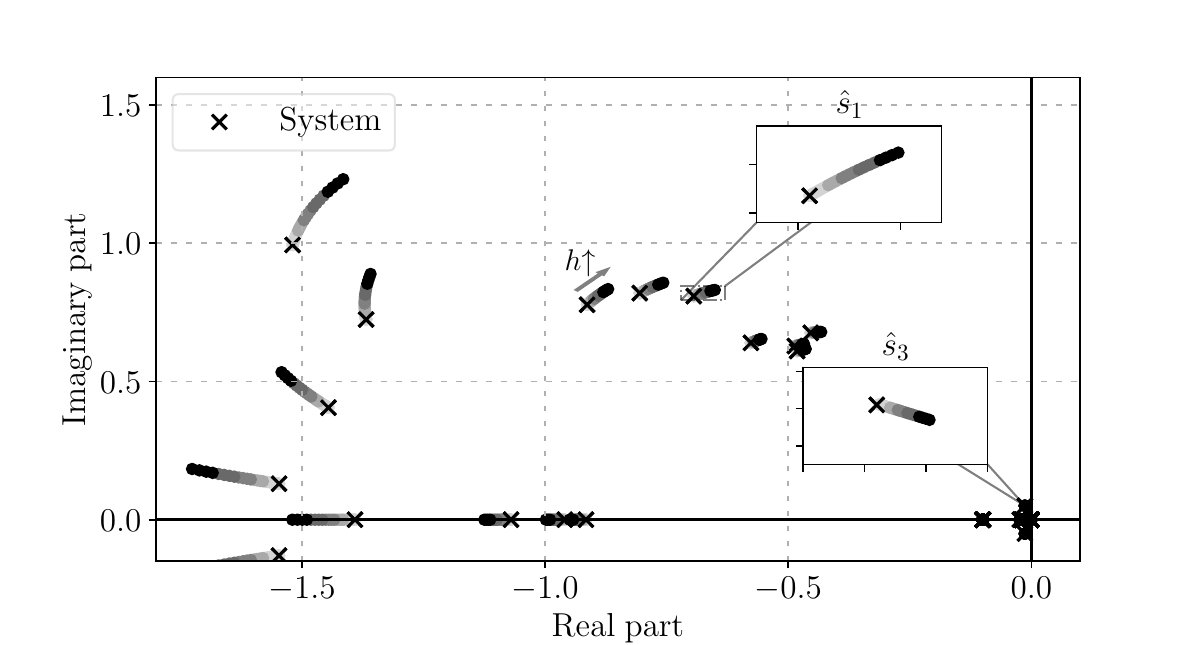}
    \caption{Rightmost eigenvalues.}
    \label{fig:ND_TMa}
  \end{subfigure}
  \hfill
  \begin{subfigure}{0.49\columnwidth}
    \centering  
    \includegraphics[width=\linewidth]{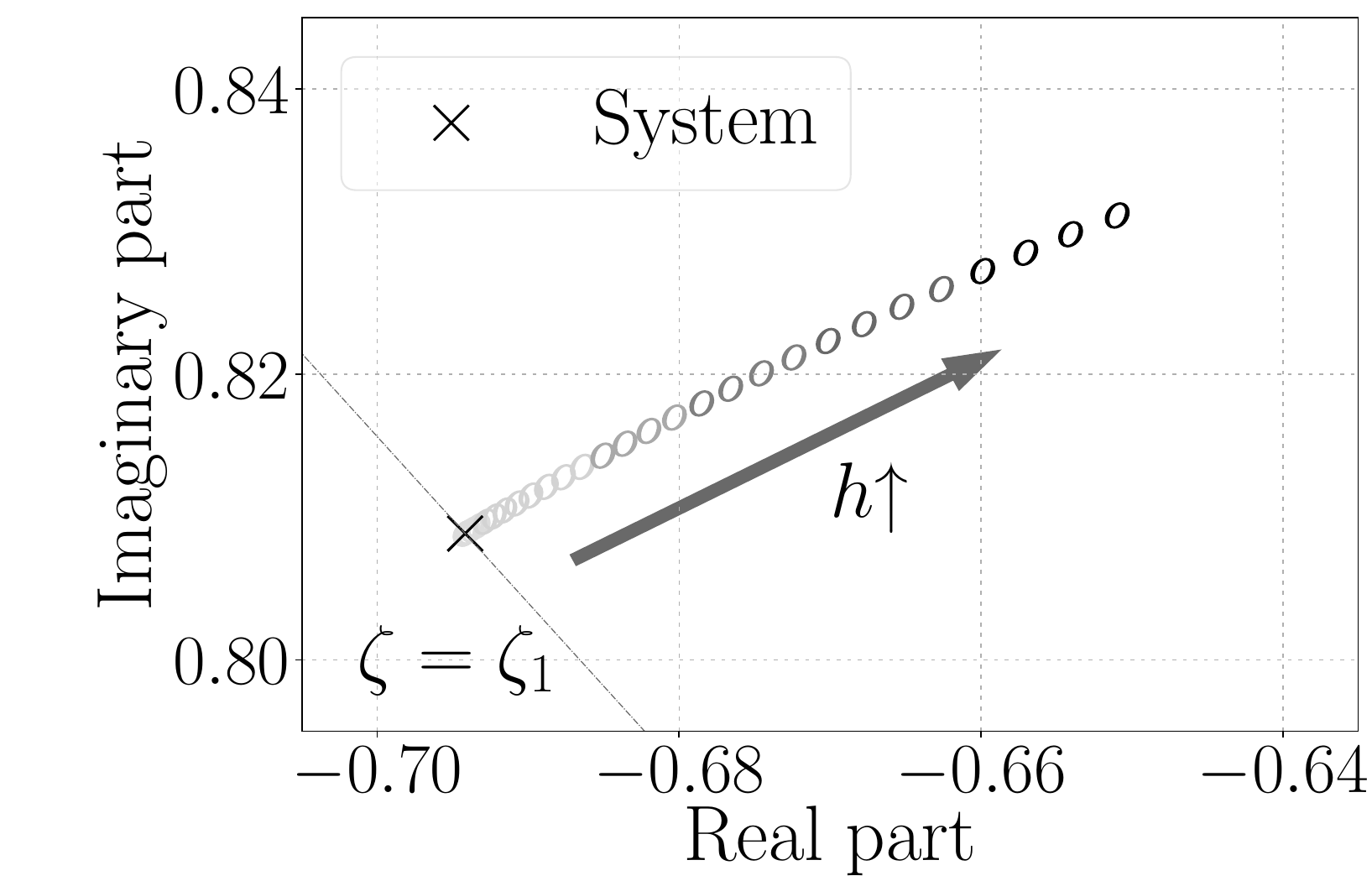}
    \caption{Close-up for eigenvalue $\hat{s}_1$.}
    \label{fig:ND_TMb}
  \end{subfigure}
  \begin{subfigure}{0.49\columnwidth}
    \centering  
    \includegraphics[width=\linewidth]{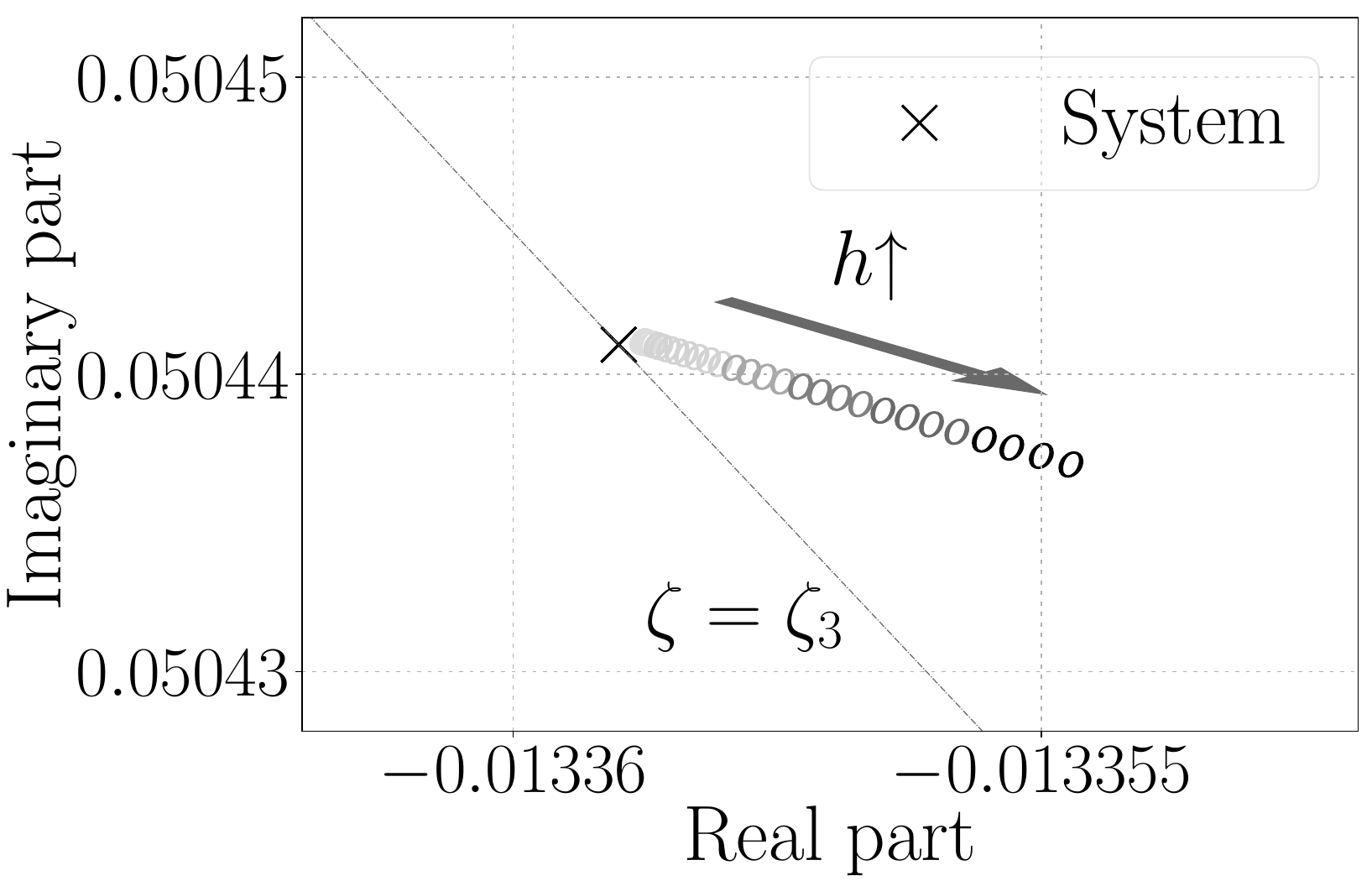}
    \caption{Close-up for eigenvalue $\hat{s}_3$.}
    \label{fig:ND_TMc}
  \end{subfigure}
  \caption{Delay-free system: Eigenvalue deformation under the effect of the \ac{tm} ($\theta=0.5$).
  }
  \label{fig:ND_TM}
\end{figure}

We illustrate how the Theta method deforms numerically the dynamic modes of the system
as the integration time step varies in the range $h \in [0.001,0.5]$~s.  In particular, Fig.~\ref{fig:ND_TMa} shows the deformation of the most critical system modes (represented by the rightmost eigenvalues) for $\theta=0.5$ (\ac{tm}).  To track the positions of the numerically deformed eigenvalues and ensure they are ordered in a consistent way, we check both the Euclidean distances of eigenvalues as well as the variations of the associated modal participation factors.   
Figures~\ref{fig:ND_TMb}--\ref{fig:ND_TMc} 
show a close-up of 
${s}_1 = -0.694176+j0.808851$ and ${s}_3 = -0.013359+j0.050441$, with the corresponding deformed eigenvalues being denoted with $\hat{s}_1$ and $\hat{s}_3$, respectively.  
The plots indicate that the \ac{tm} is very accurate to approximate the system dynamics for small time steps, whereas for very large steps it introduces spurious underdamping, yet without compromising numerical stability.  These results are as expected for a delay-free system.


\begin{figure}[ht!]
  \centering
   \begin{subfigure}{0.49\columnwidth}
    \centering  \includegraphics[width=\linewidth]{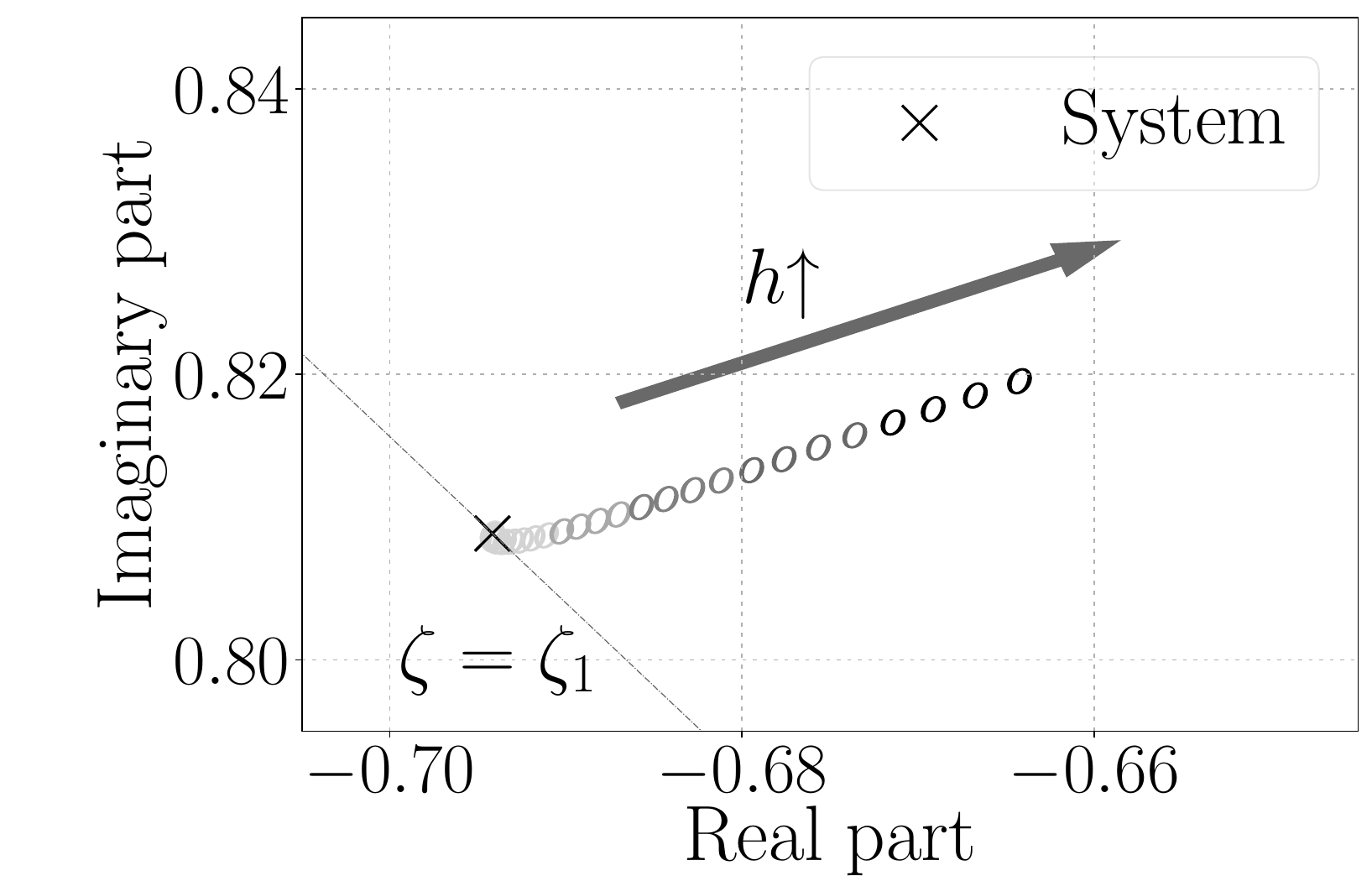}
    \caption{Close-up for eigenvalue $\hat{s}_1$.}
    \label{fig:ND_049b}
  \end{subfigure}
  \begin{subfigure}{0.49\columnwidth}
    \centering  \includegraphics[width=\linewidth]{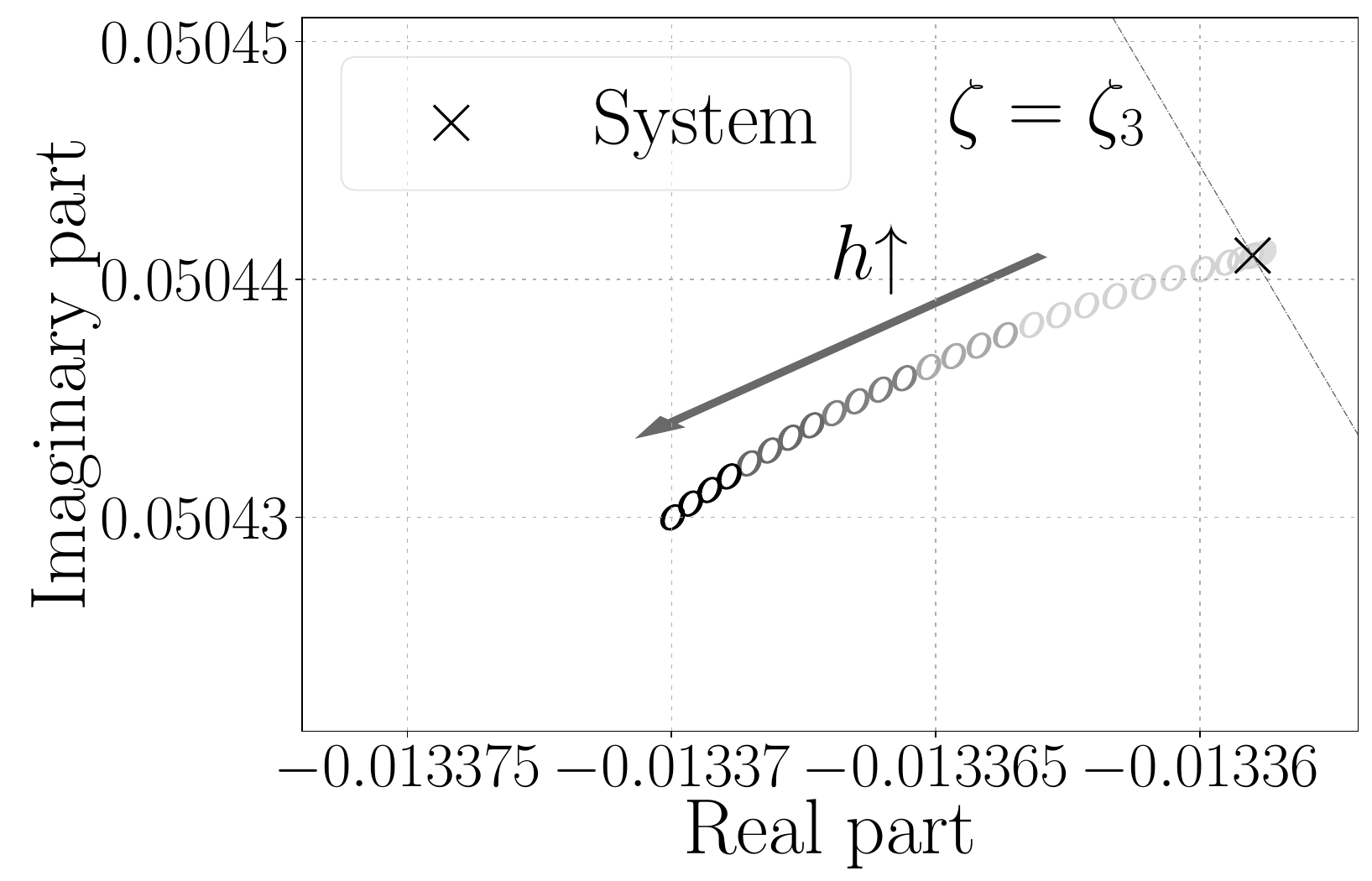}
    \caption{Close-up for eigenvalue $\hat{s}_3$.}
    \label{fig:ND_049c}
  \end{subfigure}
  \caption{Delay-free system: Eigenvalue deformation under the effect of the Theta method, for 
  $\theta=0.49$.}
  \label{fig:ND_049_sssa}
\end{figure}

\begin{figure}[ht!]
  \centering
   \includegraphics[width=\linewidth]{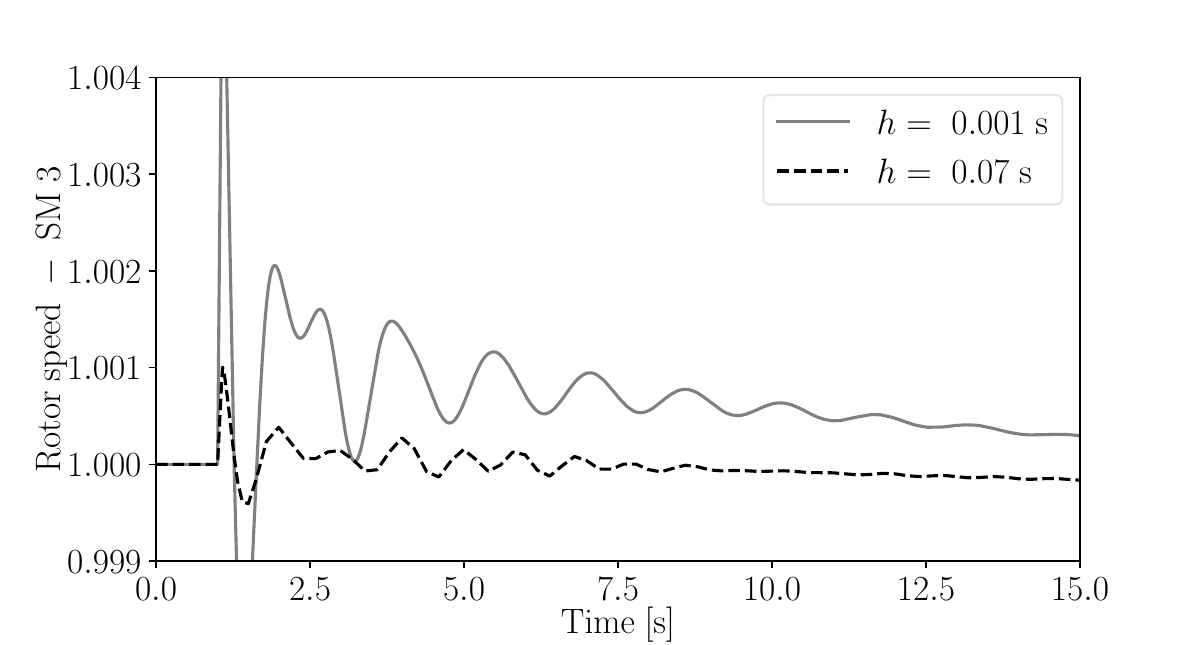}
  \caption{Delay-free system: Rotor speed of \ac{sm}~3, $\theta=0.49$. 
  }
  \label{fig:ND_049_tds_syn3}
\end{figure}

Decreasing the parameter $\theta$ to values smaller than $0.5$ 
leads to numerical methods with higher damping.  For example, setting $\theta=0$ corresponds to the \ac{bem}, which strongly overdamps all system modes.  However, not all modes are ensured to be overdamped for $0 < \theta < 0.5$.  This is especially true for values close to $0.5$, which is also a common choice in commercial software implementations  \cite{powerfactory}. 
For example, setting $\theta=0.49$ introduces a dichotomy in the response of the Theta method, where most modes get numerically underdamped as $h$ increases (like $s_1$ in Fig.~\ref{fig:ND_049b})
yet there also exist modes that appear slightly overdamped 
(e.g.,~$s_3$ in Fig.~\ref{fig:ND_049c}).
To further confirm the validity of these results, we carry out a \ac{tdi} of the nonlinear system model considering a three-phase fault at bus~6 occurring at $t=1.0$~s and cleared after $80$~ms by opening the line that connects buses 5 and~6.  
Figure~\ref{fig:ND_049_tds_syn3} shows the rotor speed of \ac{sm}~3, 
which is the variable associated to $s_{3,4}$ with the largest participation factor.  
The figure indeed indicates that the damping of the obtained oscillation is higher for larger time steps.
In Fig.~\ref{fig:ND_049_tds_syn3}, we estimate the numerical error accumulation over time using a reference trajectory obtained with  
$h = 0.0001$~s and for a duration of 15~s. 
 For $h = 0.001$~s, the maximum and average errors are $7.4 \cdot 10^{-6}$ and $1.2 \cdot 10^{-7}$, respectively, while for $h = 0.07$~s, they are $3.7 \cdot 10^{-4}$ and $-1.1 \cdot 10^{-6}$ respectively. 
%


\subsection{System with Inclusion of Delays}

We next examine the numerical deformation caused by the Theta method to a power system model that includes time delays.  To this end, we assume that the input signals of all \acp{pss}, as well as the frequency signal of the \ac{agc}, are impacted by a constant delay of $\tau=60$~ms.  The resulting model is small-signal stable as all finite eigenvalues lie in the open left half of the \textit{s}-plane.

We examine the numerical deformation introduced by the \ac{tdi} method. In particular, we begin with the \ac{tm}, 
the small-disturbance dynamics of which are assessed by finding the eigenvalues of the matrix pencil \eqref{eq:theta:FG_pencil} for $\theta=0.5$.
Conclusions are similar to the ones obtained for the delay-free system, i.e.,~the \ac{tm} remains stable for realistic time step sizes, while for large steps all modes are numerically underdamped.
We then study the effect of varying the parameter $\theta$.  To this end, we determine for different step sizes the value $\theta=\theta_{\zeta}$ for which the damping ratio $\hat \zeta$ of the deformed eigenvalue $\hat s$ closely matches the damping ratio $\zeta$ of the corresponding system eigenvalue $s$.  
The results for the pair $s_{3,4} = -0.013322 \pm j0.050422$
are shown in Table~\ref{Table_1}.  For $h=0.01$~s, the mode's damping is most accurately captured for $\theta=0.484$.  As $h$ increases, $\theta_{\zeta}$ approaches $0.5$.

\begin{table}[ht!]
\renewcommand{\arraystretch}{1.5}
\centering
\caption{System with a delay $\tau=60$ ms: Value of $\theta_{\zeta}$ for the pair of eigenvalues $\hat{s}_{3,4}$, for different values of $h$.}
 \begin{threeparttable}
\begin{tabular} { c | ccccccccc}
    \toprule\toprule
    $h$~[s] & $0.01$ 
    & $0.03$ & $0.06$ & $0.08$ & $0.1$ 
    & $0.15$ & $0.2$ \\
    \midrule
    $\theta_{\zeta}$ & $0.484$ 
    & $0.494$ & $0.497$ & $0.498$ & $0.498$ 
    & $0.499$ & $0.499$ \\
    \bottomrule \bottomrule
\end{tabular}
\end{threeparttable}
\label{Table_1}
\end{table}

\begin{figure}[ht!]
  \centering
  \begin{subfigure}{\columnwidth}
    \centering
    \includegraphics[width=\linewidth]{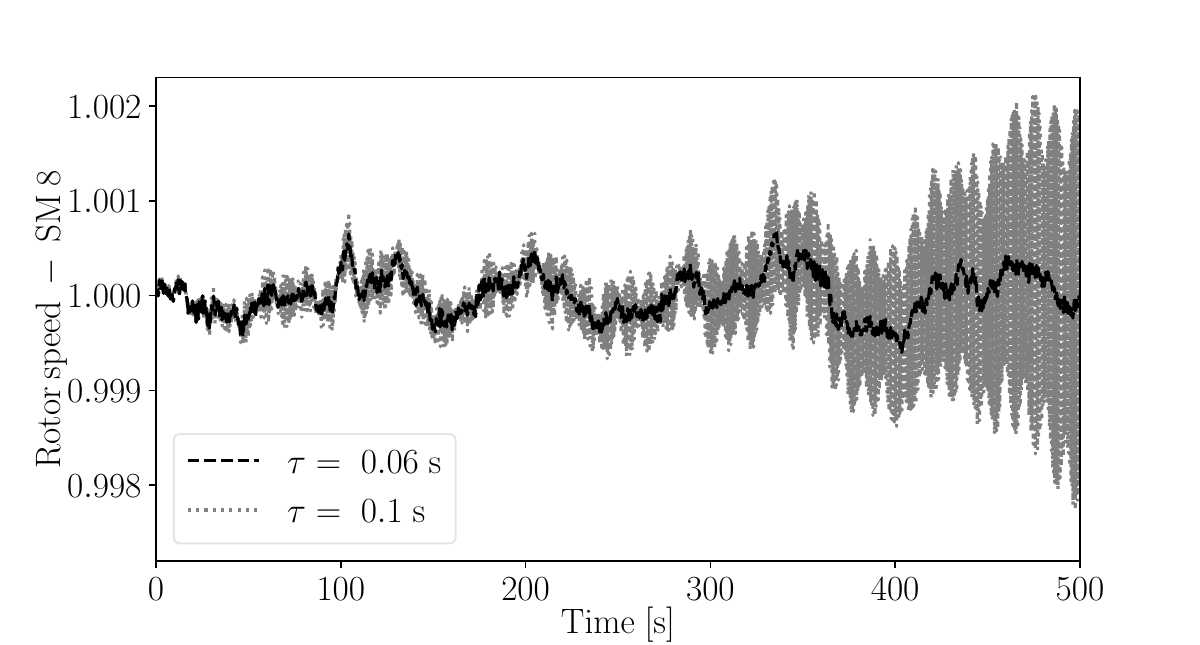}
    \caption{Effect of increasing $\tau$ from $60$ to $100$~ms, $h=0.001$~s.}    \label{fig:TDS_instab_stochastic_var_d_60vs100ms}
  \end{subfigure}
  \hfill
  \begin{subfigure}{\columnwidth}
    \centering
    \includegraphics[width=\linewidth]{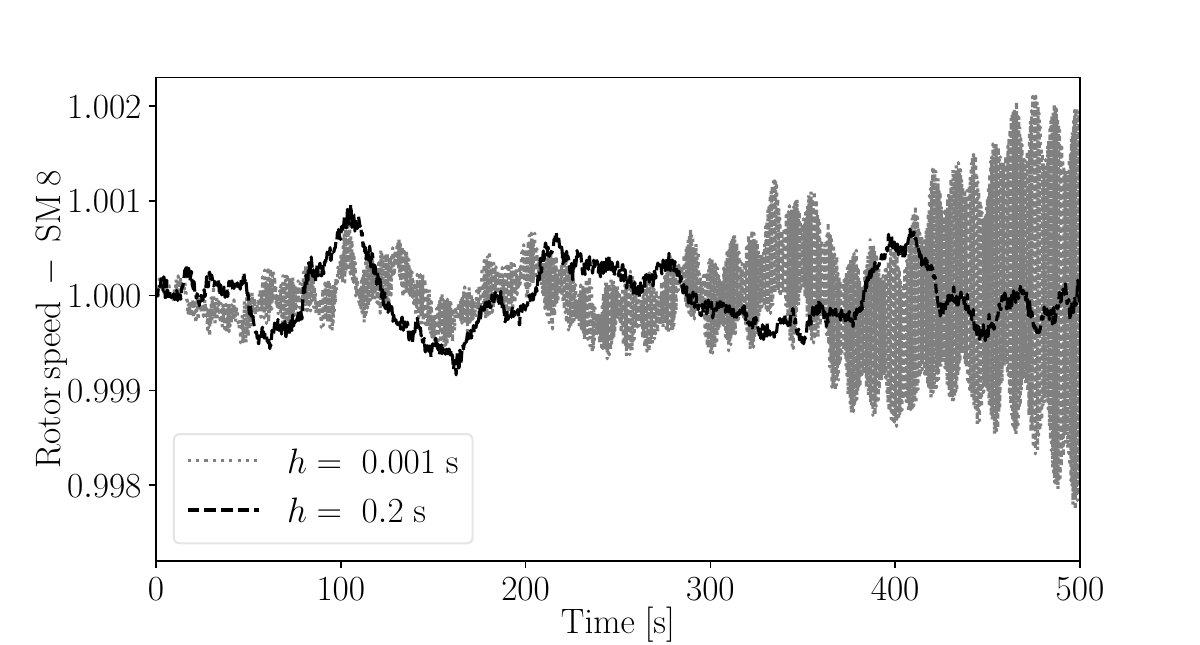}
    \caption{ Effect of increasing $h$ from $0.001$~s to $0.2$~s, $\tau = 100$~ms.} 
    \label{fig:TDS_instab_stochastic_var_d}
  \end{subfigure}
  \caption{Rotor speed of \ac{sm}~8 in the presence of stochastic loads.}
  \label{fig:noise_unstable}
\end{figure}

By increasing the value of the time delay $\tau$ from 60~ms to 100~ms, a complex pair of eigenvalues moves to the right half of the \textit{s}-plane and the system becomes unstable.
This is consistent with Fig.~\ref{fig:TDS_instab_stochastic_var_d_60vs100ms}, which shows for $h=0.001$~s the time-domain response of the rotor speed of SM~8 assuming that the active power consumption of the load connected to bus~16 is stochastic.

\begin{figure}[ht!]
  \centering
  \begin{subfigure}{\columnwidth}
    \centering
    \includegraphics[width=\linewidth]{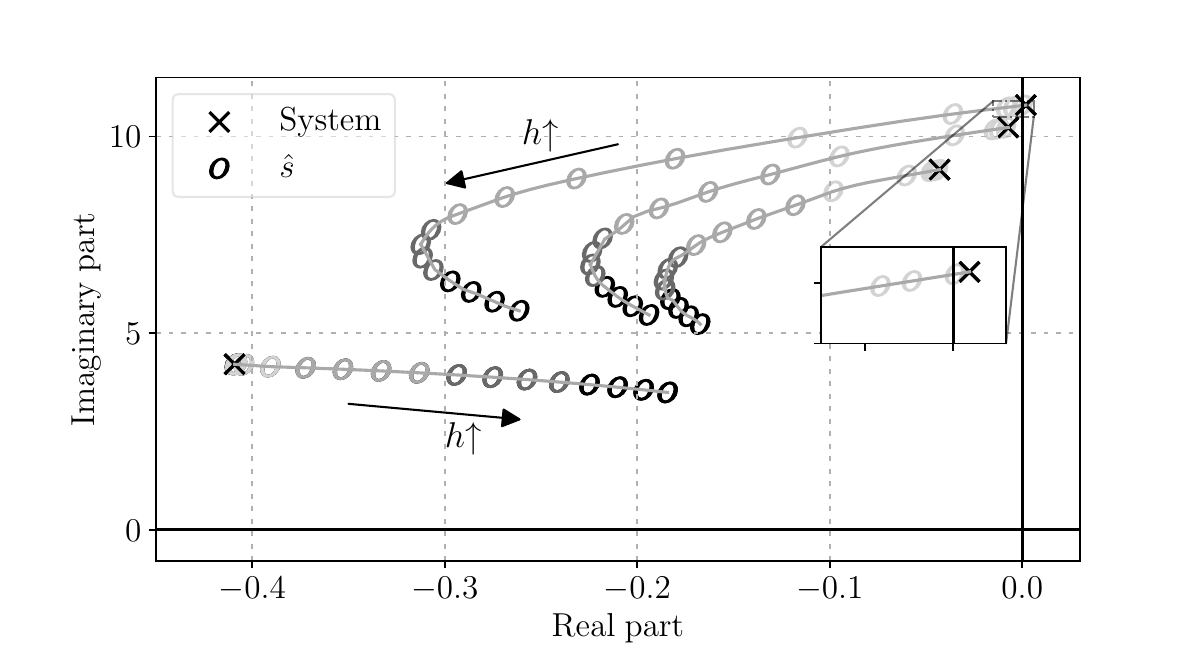}
    \caption{Eigenvalue deformation as $h$ is varied. The system is small-signal unstable.}
    \label{fig:instab_d01_TM}
  \end{subfigure}
  \hfill
  \begin{subfigure}{\columnwidth}
    \centering
    \includegraphics[width=\linewidth]{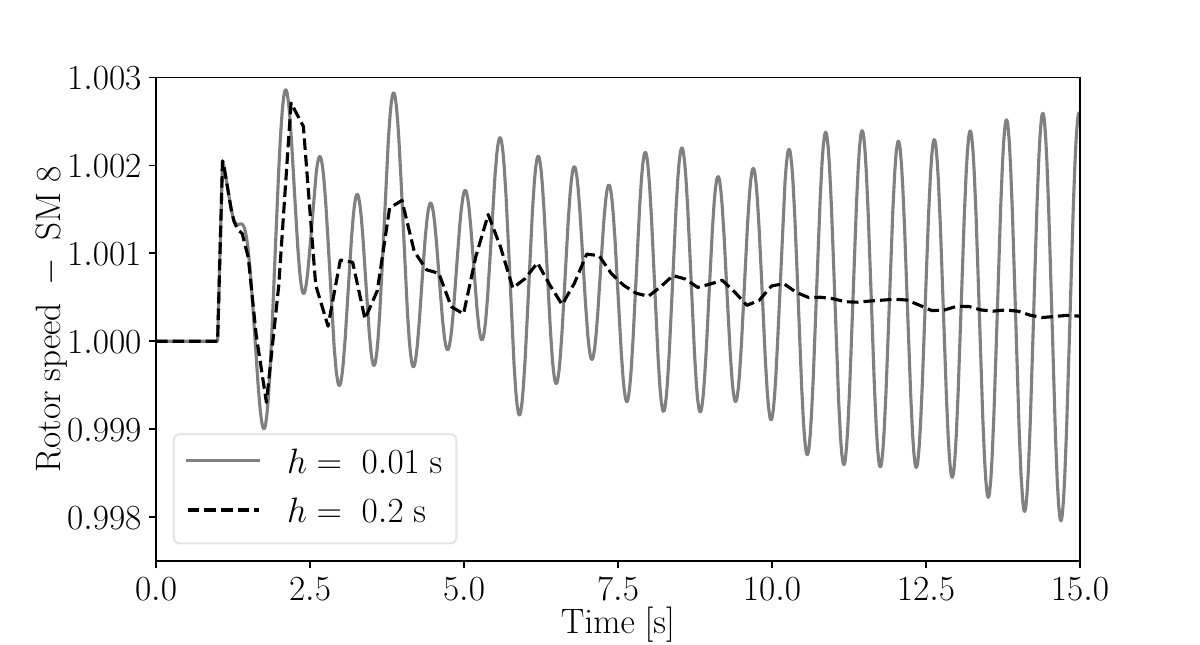}
    \caption{Rotor speed of \ac{sm}~8 following a three-phase fault.} 
    \label{fig:TDS_instab_d01_TM_ALT}
  \end{subfigure}
  \caption{System with $\tau=100$~ms solved with \ac{tm} ($\theta = 0.5$).}
  \label{fig:unstable}
\end{figure}

When we apply the \ac{tm} in this case, Fig.~\ref{fig:instab_d01_TM} shows that with an increase of $h$ beyond $0.01$~s, the method makes the system appear stable, despite actually being unstable.
This result is consistent with  
Fig.~\ref{fig:TDS_instab_stochastic_var_d}, which indicates that for $h = 0.2$~s, the numerical method fails to capture the instability of the system.
%
For completeness, we further check this result for the case of a large disturbance, by considering the same contingency as above, i.e.,~a three-phase fault at bus~6.  
Figure~\ref{fig:TDS_instab_d01_TM_ALT} shows the rotor speed of \ac{sm}~8 following the contingency.
The maximum and average numerical errors with (respect to a reference trajectory obtained with $h = 0.0001$~s) are, respectively, $1.9 \cdot 10^{-3}$ and $-5.9 \cdot 10^{-6}$ for $h = 0.01$~s; and $1.7 \cdot 10^{-3}$ and $-2.0 \cdot 10^{-5}$ for $h = 0.2$~s.
The plot confirms consistency with the \ac{sssa} results for a large enough time step ($h=0.2$~s).

\begin{figure}[ht!]
  \centering
  \centering
  \includegraphics[width=\linewidth]{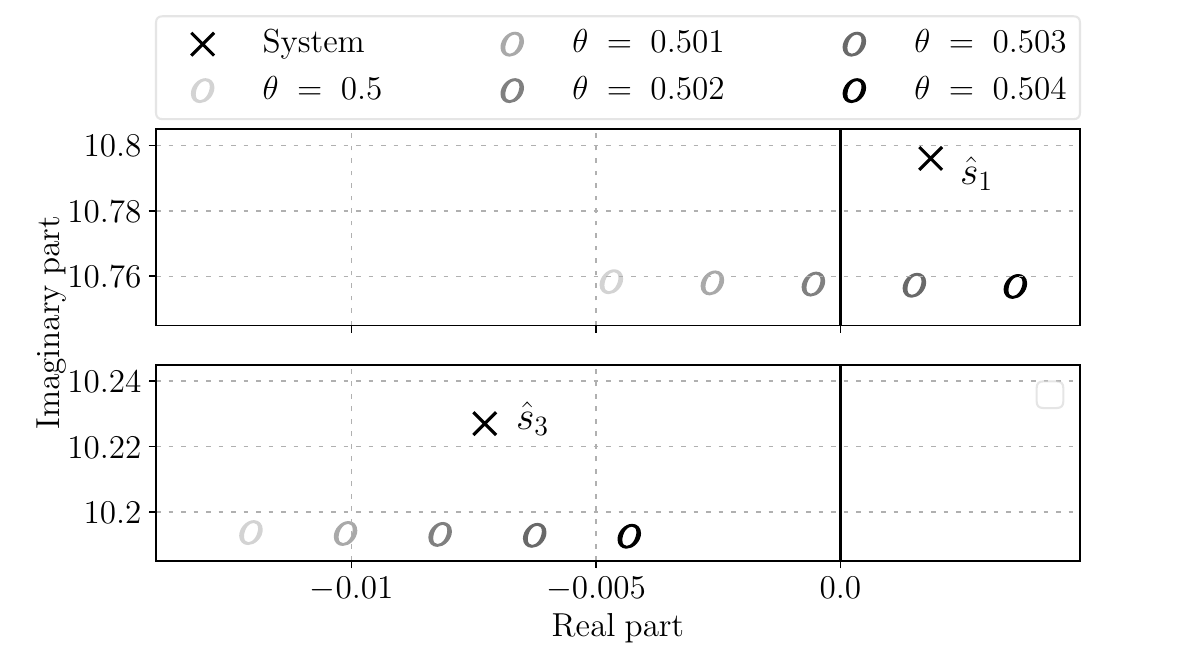}
  \caption{System with delays: Effect of $\theta$ on the deformation of the two rightmost modes, $h=0.02$~s.}
  \label{fig:instab_d01_theta_002}
\end{figure}

\begin{figure}[ht!]
  \centering
  \begin{subfigure}{0.49\columnwidth}
    \centering
    \includegraphics[width=\linewidth]{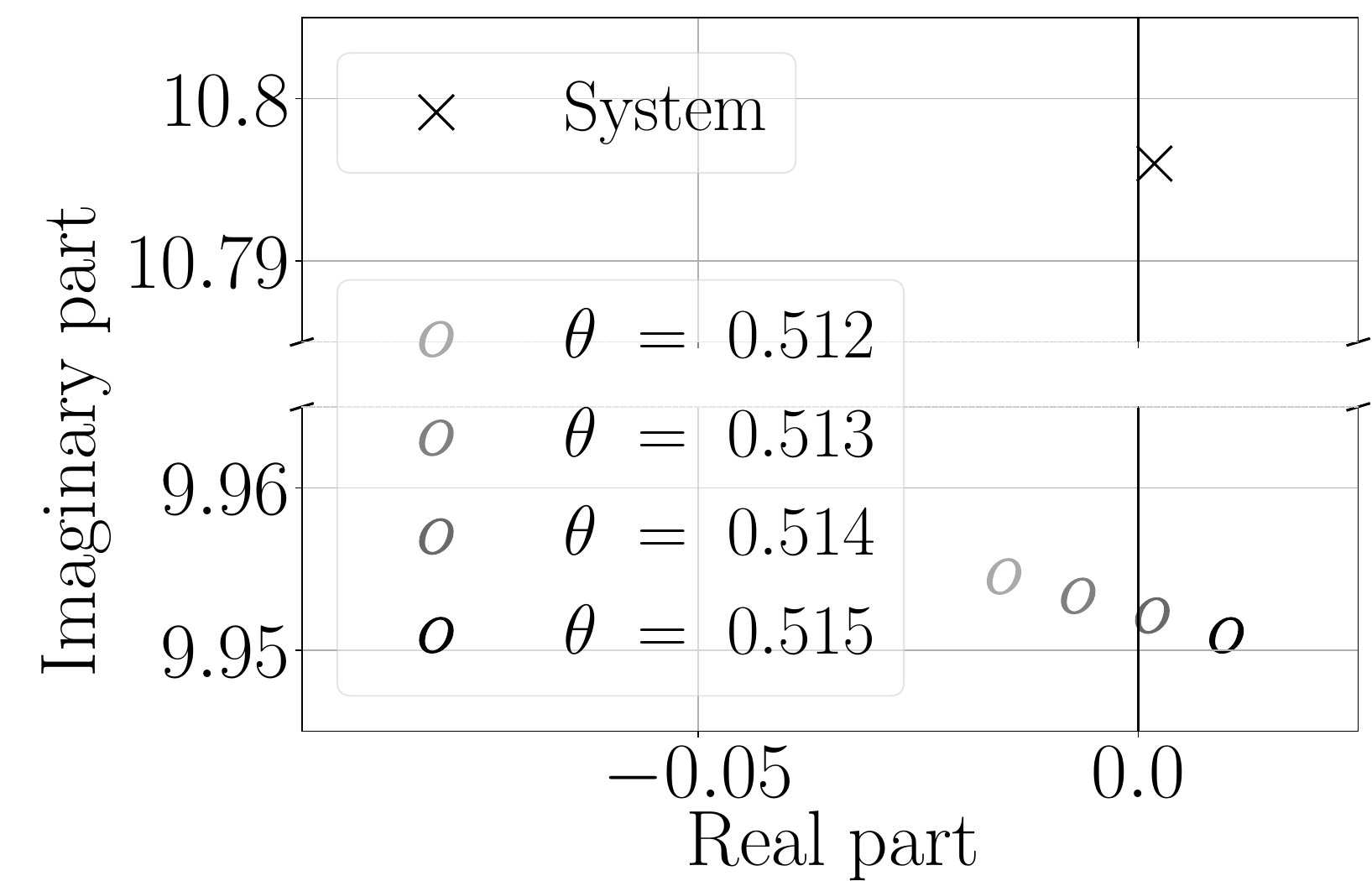}
    \caption{$\hat{s}_{1}$.}
    \label{fig:instab_d01_theta_h01a}
  \end{subfigure}
  \hfill
  \begin{subfigure}{0.49\columnwidth}
    \centering
    \includegraphics[width=\linewidth]{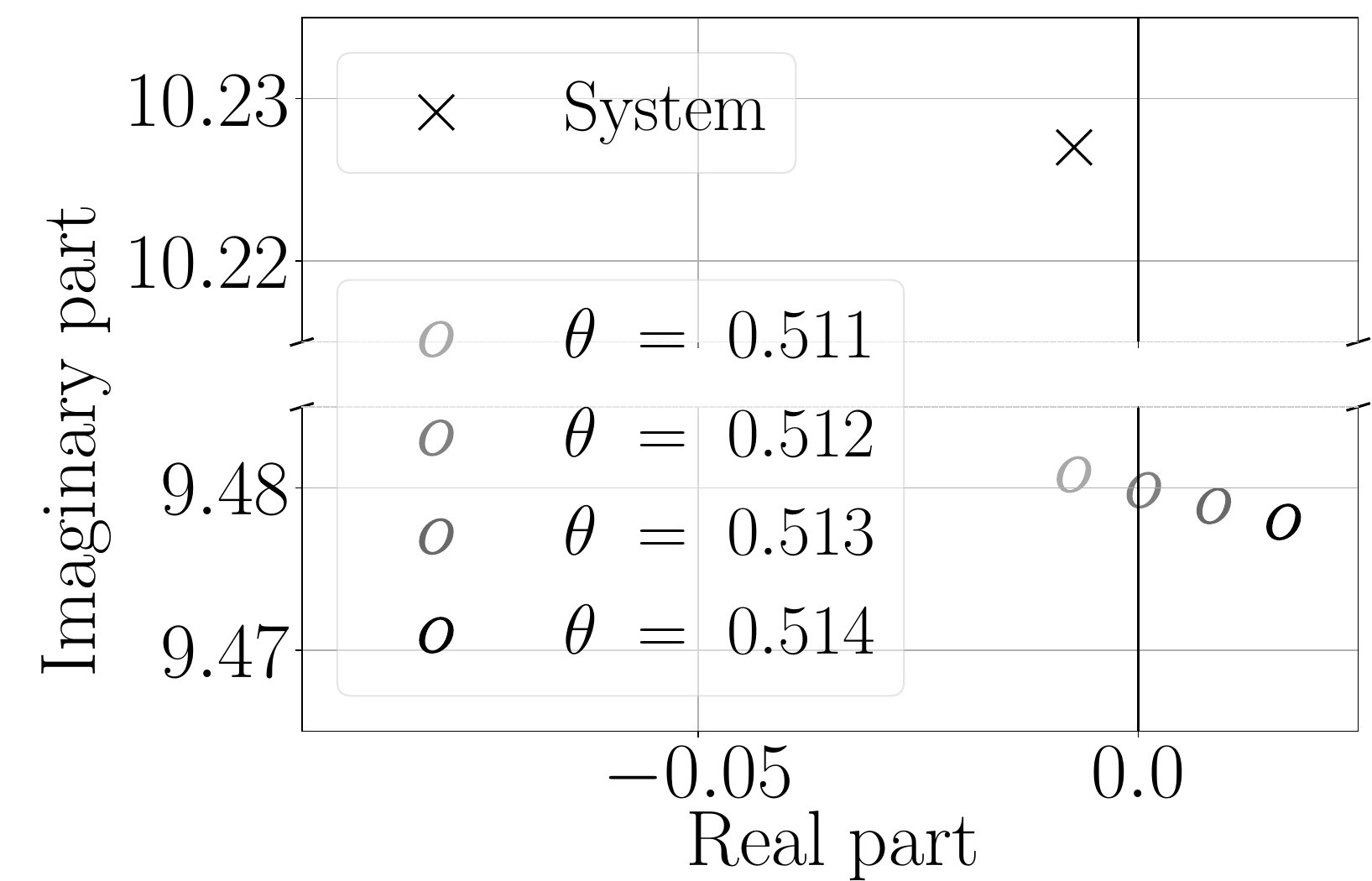}
    \caption{$\hat{s}_{3}$.}
    \label{fig:instab_d01_theta_h01b}
  \end{subfigure}
  \caption{System with delays: Effect of $\theta$ on the deformation of the two rightmost modes, $h=0.1$~s.}
  \label{fig:instab_d01_theta_h01}
\end{figure}

We then study the effect of the $\theta$ parameter on the precision of the \ac{tdi} method. 
To this end, we examine the deformation caused by the Theta method to the two rightmost pairs of eigenvalues.
We observe that for small time steps a slight variation of $\theta$ allows to capture the instability of the system.  This is illustrated in 
Fig.~\ref{fig:instab_d01_theta_002} where, 
$\theta=0.503$ leads to consistent results between \ac{sssa} of the system and the Theta method.
Nevertheless, there are practical restrictions for large steps.  In particular, attempting to accurately represent one mode through variation of $\theta$ may negatively impact the accuracy of another mode.  This is because faster dynamics typically suffer from larger numerical distortions under the same variation of $h$, with the phenomenon being more prominent for large steps.  An example of such a behavior is shown in Fig.~\ref{fig:instab_d01_theta_h01}.

\subsection{Modified System with 
Inverter-Based DERs}
\label{sec:case:modified}

This section presents simulation results based on a modified version of the IEEE 39-bus system, where part of the synchronous generation has been replaced by inverter-based \acp{der}.  In particular, the \acp{sm} at buses 30, 34, 35, 36 and 37 have been replaced by \acp{der} of the same capacity.  Each \ac{der} comprises an inner control loop that regulates the $d$ and $q$ components of the current in the $dq$ reference frame, and two outer loops for primary frequency and voltage control, respectively
\cite{ORTEGA201837}.  The input signal of the frequency control of each \ac{der} is provided by a \ac{srf-pll}. 

\begin{figure}[ht!]
    \centering
    \includegraphics[width=\linewidth]{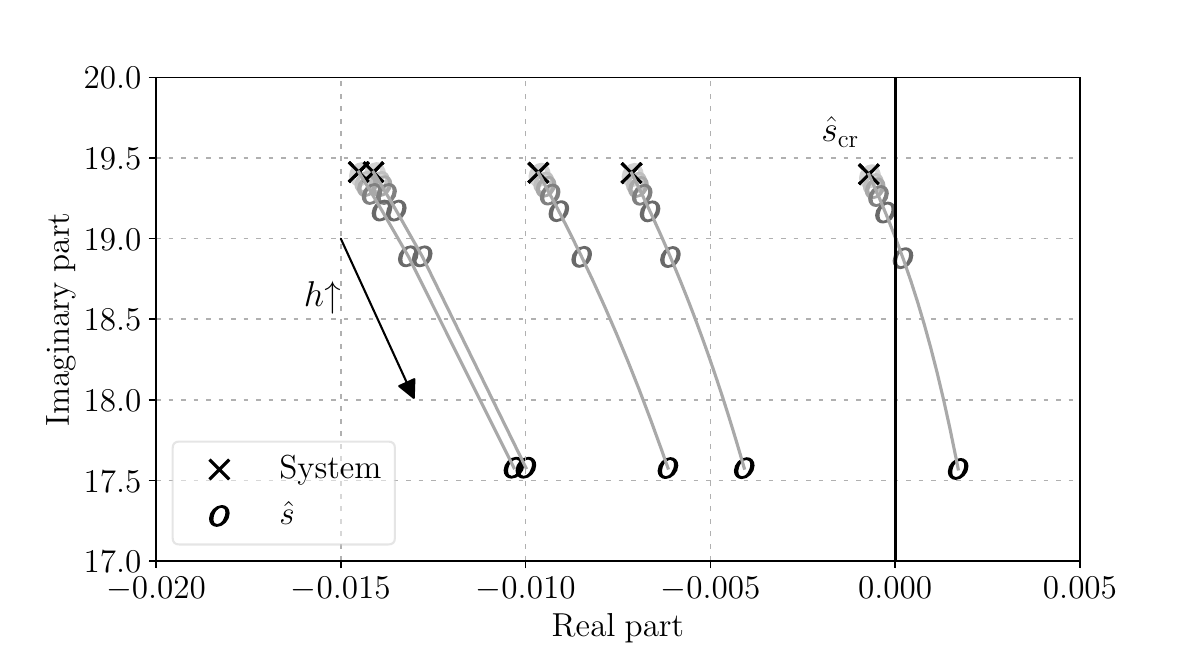}
    \caption{Modified system with delays: Deformation of rightmost eigenvalues under the effect of \ac{tm}.}
  \label{fig:mod39b_stab_d006_TM}
\end{figure}

To study the impact of time delays, 
we assume that the measurements of the \ac{der} voltage and frequency controllers, the \acp{pll}, the \acp{pss} and the \ac{agc} are all impacted by a $60$~ms delay.
For the needs of this example, 
we make the \acp{pll}
oscillatory, a behavior often observed 
under weak-grid conditions \cite{PLL_3_Avila}.  
Figure~\ref{fig:mod39b_stab_d006_TM} depicts the rightmost eigenvalues of the modified system, as well as the corresponding numerically deformed eigenvalues produced by the \ac{tm}. 
As it can be seen, despite the system being stable, 
increasing $h$ to large enough values makes it appear unstable, as the real part of 
a deformed eigenvalue ($\hat{s}_{\rm cr}= -0.000711+j19.399$) becomes positive. 
In other words, the \ac{tm} is numerically unstable in this scenario. 


\begin{figure}[ht!]
    \centering
    \includegraphics[width=\linewidth]{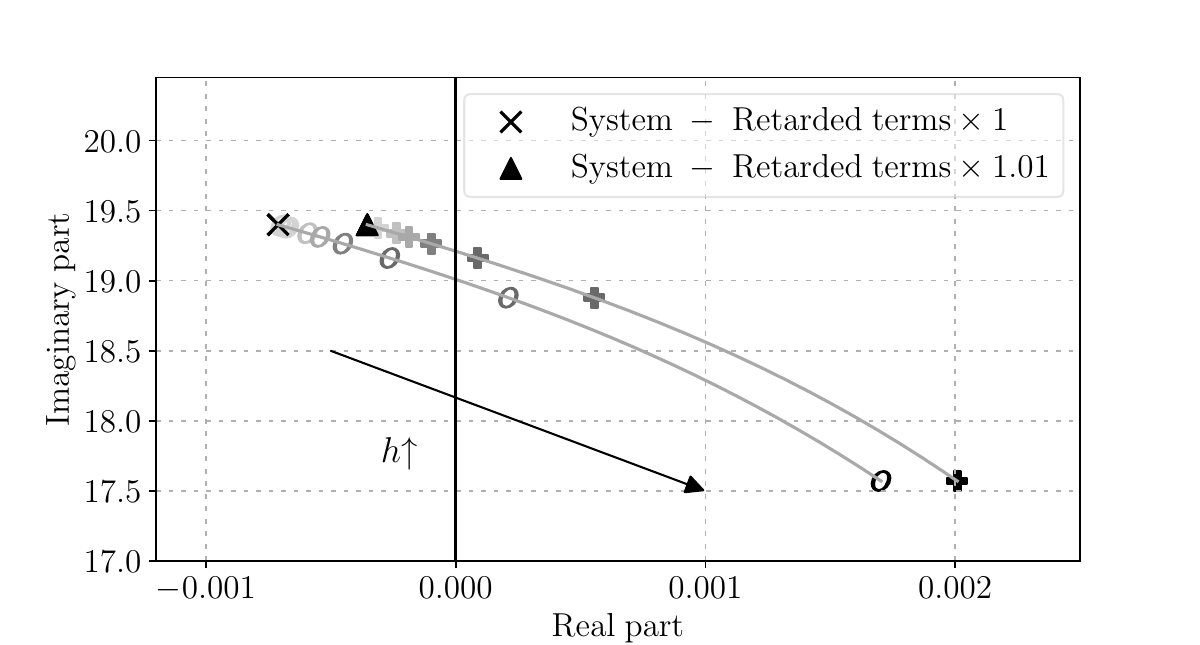}
    \caption{Modified system with delays: Deformation of rightmost eigenvalues caused by increased magnitudes of coefficients of delayed variables, 
    \ac{tm}. The \ac{tm} is destabilized for $h\geq0.02$~s.}
  \label{fig:mod39b_stab_d006_TM_b101}
\end{figure}

We next study the effect of  
varying the magnitudes of the coefficients of delayed terms in the system's equations, on numerical stability.  Results 
are presented in Fig.~\ref{fig:mod39b_stab_d006_TM_b101} and indicate that even a slight variation of $1\%$ significantly impacts the stability of the \ac{tdi} method.  In particular, a time step of $h=0.02$~s is enough to destabilize the \ac{tm} in this scenario.
 

\begin{figure}[ht!]
  \centering
   \begin{subfigure}[t]{0.49\columnwidth}
    \centering  \includegraphics[width=\linewidth]{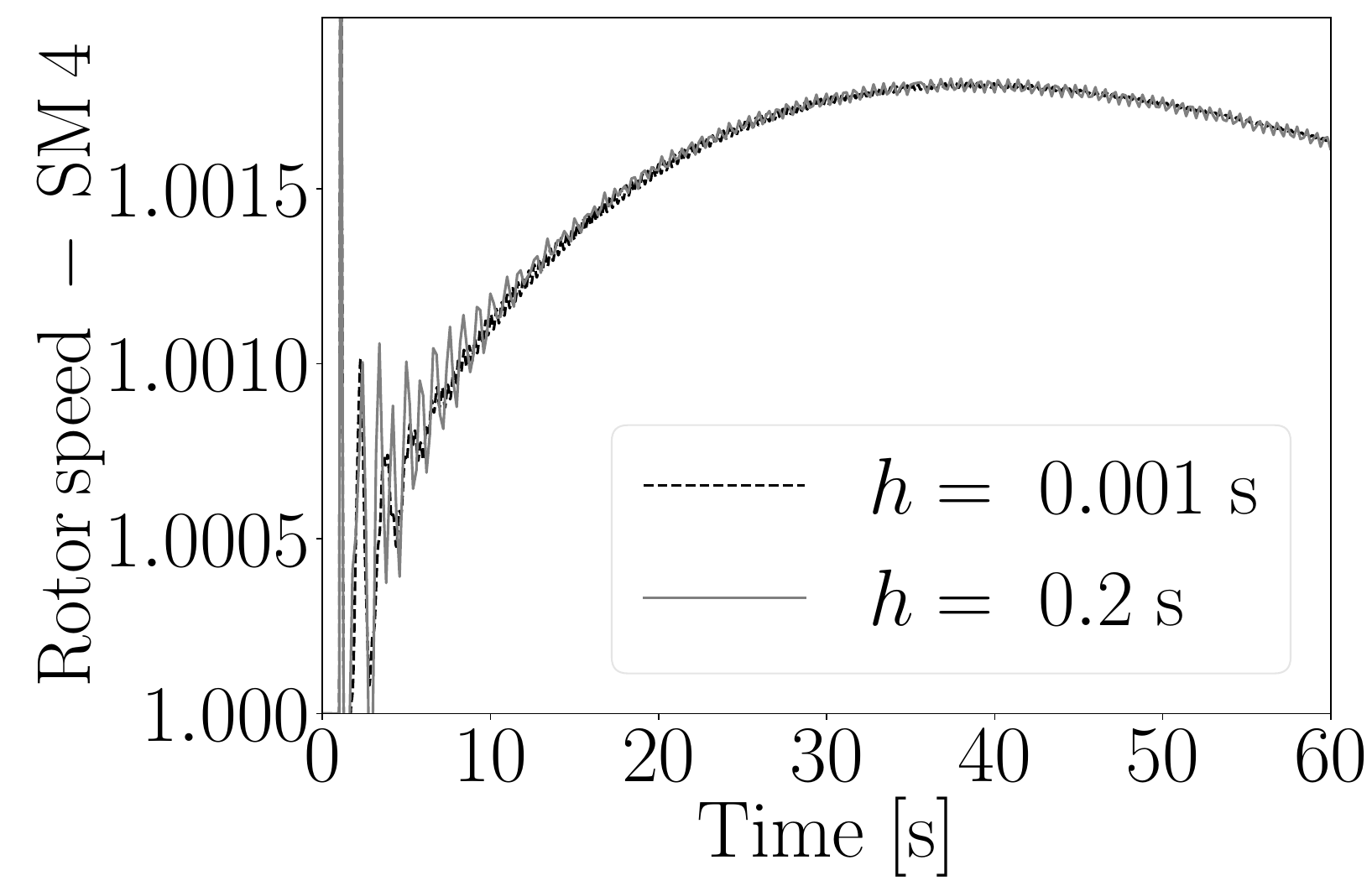}
    \caption{Without delays.}
    \label{fig:mod_d006_tds_plain}
   \end{subfigure}
   \begin{subfigure}[t]{0.49\columnwidth}
    \centering  \includegraphics[width=\linewidth]{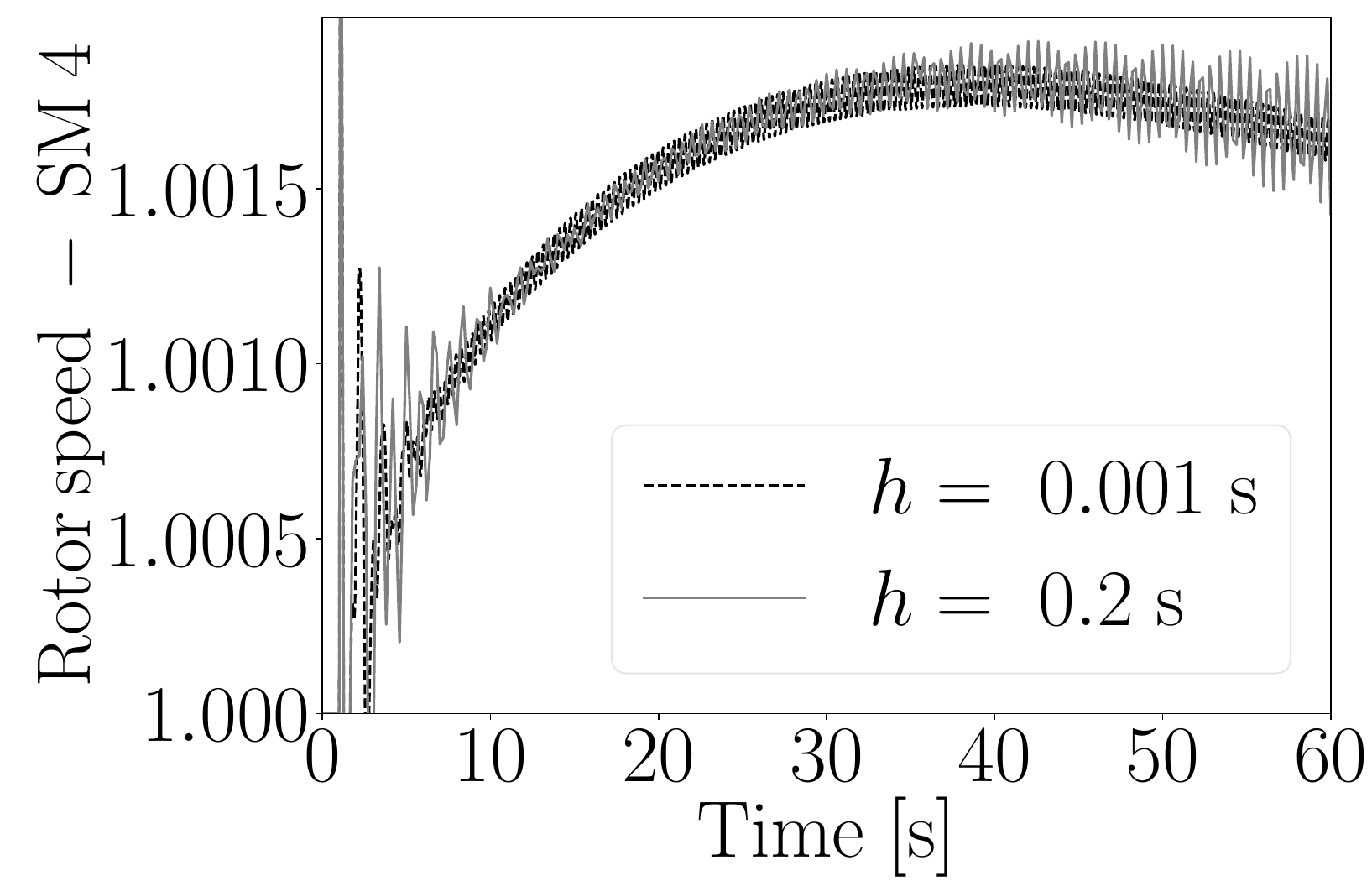}
    \caption{With delays.}
    \label{fig:mod_d006_tds_plain_del}
  \end{subfigure}
  \caption{Modified system with delays: Rotor speed of \ac{sm}~4 for $\theta=0.5$ (TM) following a three-phase fault at bus~6.}
  \label{fig:mod_d006_tds_TM}
\end{figure}

We further check the above conclusions 
through a \ac{tdi} of the system.
In particular, we consider  
a three-phase fault occurring at bus~6 at $t = 1$~s and cleared after $80$~ms by opening the line that connects buses 5 and 6.  
The response of the rotor speed of \ac{sm}~4
is displayed in Fig.~\ref{fig:mod_d006_tds_TM}.
For the sake of comparison, the delay-free system is also plotted in the same figure. 
Figure~\ref{fig:mod_d006_tds_plain_del} confirms that for a large enough time step (in this case $0.2$~s), the \ac{tm} makes an otherwise stable trajectory of the system 
appear unstable.
The maximum and average numerical errors 
are $5.6 \cdot 10^{-5}$ and $-4.4 \cdot 10^{-7}$ for $h = 0.001$~s; and $7.6 \cdot 10^{-4}$ and $8.6 \cdot 10^{-6}$ for $h = 0.2$~s.

\begin{figure}[ht!]
  \centering
   \begin{subfigure}[t]{\columnwidth}
    \centering  \includegraphics[width=\linewidth]{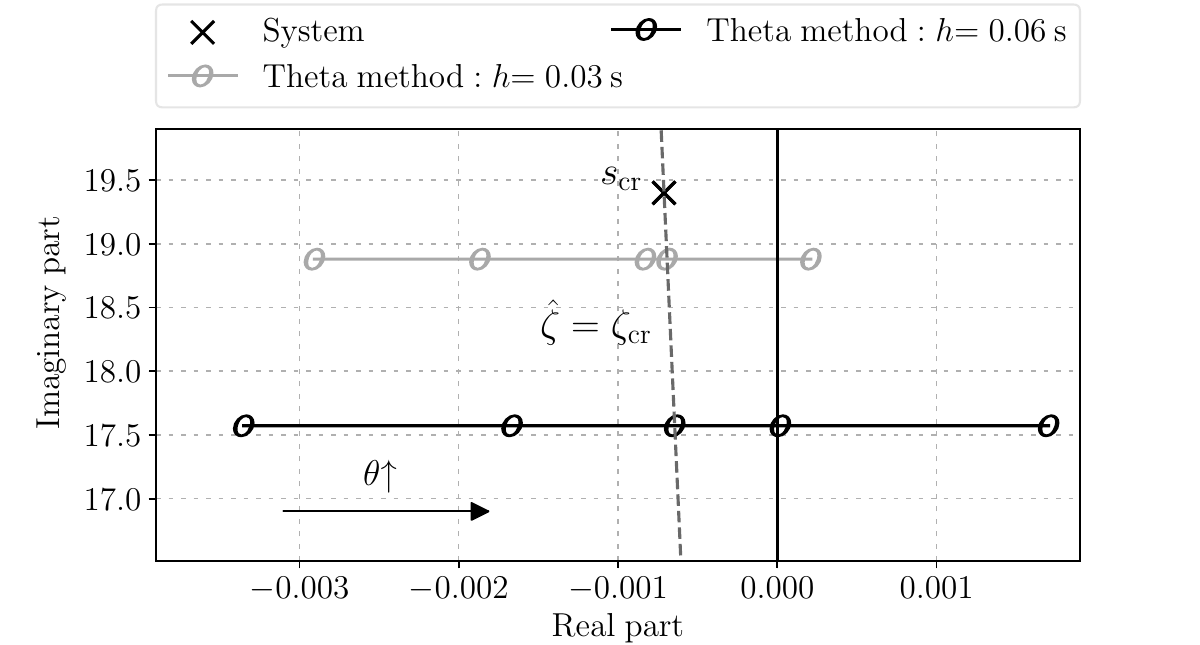}
    \caption{Effect of a very small variation of $\theta$ from $0.4997$ to $0.5$.}
    \label{fig:mod39b_stab_theta}
   \end{subfigure}
   \begin{subfigure}[t]{\columnwidth}
    \centering  \includegraphics[width=\linewidth]{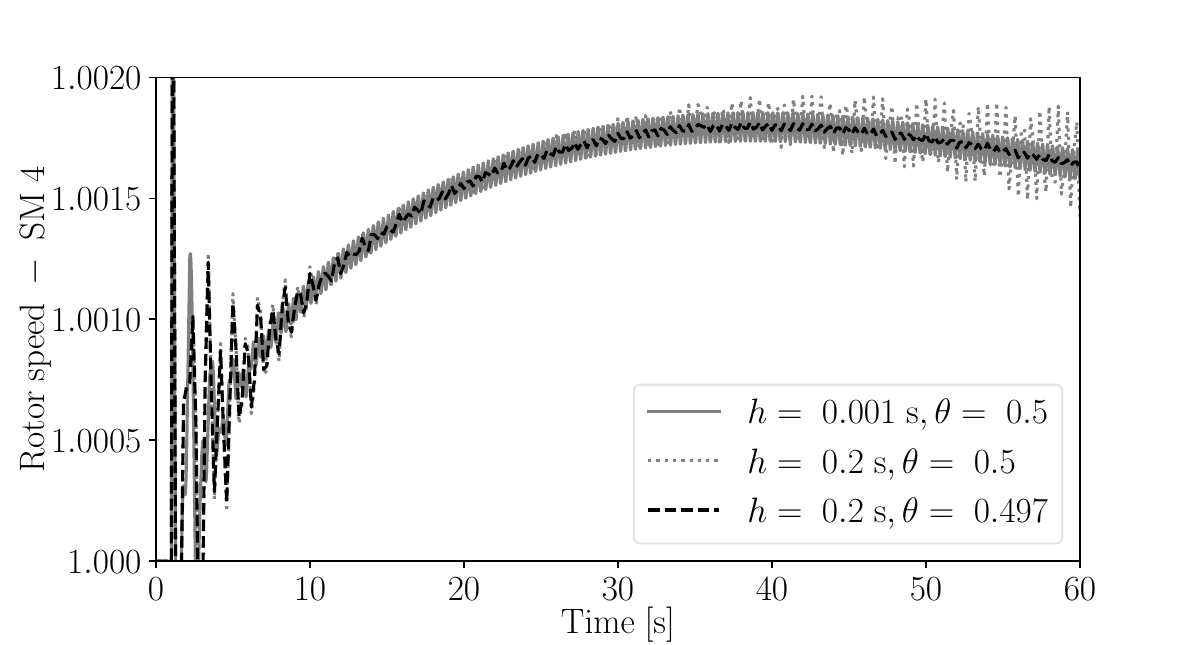}
    \caption{Rotor speed of \ac{sm}~4 following a three-phase fault at bus~6.} 
    \label{fig:mod39b_stab_theta_syn4_tds}
  \end{subfigure}
  \caption{Modified system with delays: Effect of varying $\theta$ on the precision of the Theta method.}
  \label{fig:mod_d006_tds_Theta}
\end{figure}

We finally examine how the $\theta$ parameter can be employed to compensate for the numerical deformation introduced by the Theta method in this scenario. 
In particular, $\theta$ is altered to a value $\theta_{\zeta}<0.5$ for which $\hat{s}_{\rm cr}$ has the same damping $\zeta_{cr}$ as $s_{\rm cr}$, i.e. $\hat \zeta=\zeta_{\rm cr}$.
In Fig.~\ref{fig:mod39b_stab_theta} it is shown that $\theta$ needs to be slightly modified by about $10^{-5}$ for $h=0.03$~s and 
by about $1.5 \cdot 10^{-5}$
for $h=60$~ms.
A time-domain simulation of the system further confirms this conclusion, see Fig.~\ref{fig:mod39b_stab_theta_syn4_tds}.

\section{Conclusions}
\label{sec:conclusion}

The paper shows that standard implicit \ac{tdi} methods, such as the trapezoidal and Theta method, can become numerically unstable when applied to power system models impacted by multiple time-delayed variables.  The mechanism behind this numerical phenomenon is first shown for a linear test delay differential equation. Subsequently, a systematic analysis that accounts for the dynamics of real-world power system models is carried out through a unified numerical stability and accuracy analysis based on \ac{sssa}.  Theoretical findings of the paper are fully supported by simulation results conducted based on the IEEE 39-bus system.  

The findings of this work have relevant practical applications.  Specifically, they can be employed to internally check numerical stability and accuracy of a given integration method within the time-domain simulation routines of modern power system software. The proposed matrix pencil approach enables the selection of an appropriate time step size to meet prescribed simulation accuracy criteria while accounting for the deformation of the system’s dynamics introduced by the integration method.  Naturally, reducing the time step size enhances accuracy but increases computational cost.  Additionally, these findings can aid in comparing alternative \ac{tdi} methods to achieve accurate and stable simulations without requiring a substantial reduction in time step size. 

Future work will focus on the effect of adaptive step-size algorithms, as well as on comparisons with the performance of specialized \ac{tdi} methods for delay differential equations, such as the family of Radau~IIA methods.

\appendix

\subsection[Derivation]{Derivation of \eqref{eq:theta} }
\label{deriv:theta}


Applying the Laplace transform $\mathcal{L}\{ \cdot \}$ to \eqref{eq:dae} and omitting for simplicity the initial conditions:

\begin{equation}
  \begin{aligned}
    \label{eq:laplace}
    s \mathcal L\{ {\bfg x} \} 
    &= 
    \mathcal L\{
    \bfg f(\bfg x, \bfg y)\}    \, , 
    \\
    \bfg 0_{\ny,1} &=  \mathcal L\{
    \bfg g(\bfg x, \bfg y)\}    \, .
  \end{aligned}    
\end{equation}
Using \eqref{eq:theta:ztos} and \eqref{eq:discrxy}, we obtain:
\begin{equation}
  \begin{aligned}
    \label{eq:theta:zdomain:dae}
    {(z - 1)} \mathcal Z\{ {\bfg x_n} \} 
    &= 
    h(\theta + (1-\theta) z) \mathcal Z\{
    \bfg f(\bfg x_n, \bfg y_n)\}
    \, ,     \\
    \bfg 0_{\ny,1} &=  \mathcal Z\{
    \bfg g(\bfg x_{n+1}, \bfg y_{n+1}\}  \, ,
  \end{aligned}    
\end{equation}
where $\mathcal{Z}\{ \cdot \}$ denotes the $Z$-transform.
By applying the inverse $Z$-transform to \eqref{eq:theta:zdomain:dae}, we arrive to \eqref{eq:theta}.

\subsection[Proof]{Proof of \eqref{eq:theta:FG} }
\label{proof:FG}

The proof of \eqref{eq:theta:FG} 
goes as follows:

\vspace{2mm}
\begin{proof}
Let us rewrite \eqref{eq:theta:ABC} as follows:
\begin{align*}
\bfb E \bfb x_{n+1} &= 
\bfb D_0 \bfb x_{n}
+ \bfb D_1 \bfb x_{n-1} + 
\ldots +
\bfb D_{\nd} \bfb x_{n-\nd} \, ,
\end{align*}
where 
$\bfb D_0=\bfb A + \bfb B_0$,
$\bfb D_j=\bfb B_j+\bfb C_j$, $j=1,2,\ldots,\nd-1$, and 
$\bfb D_\nd = \bfb C_\nd $.
Then, by setting:
\begin{align*}
  \bfb y_{n+1}^{[i]} &= \bfb x_{n+1-i}  \, , 
  \ \ \ \
   \bfb y_{n}^{[i]} = \bfb x_{n-i} \, ,
\end{align*}
where $i=0,1,2,\ldots,\nd$, we have equivalently:
\begin{align*}
\bfb M \bfb y_{n+1}^{[0]} &= 
\bfb D_0 \bfb y_{n+1}^{[1]} + 
\bfb D_1 \bfb y_{n+1}^{[2]} + 
\hspace{-0.5mm}
\ldots 
\hspace{-0.5mm} +
\bfb D_{\nd-1} \bfb y_{n+1}^{[\nd]} +
\bfb D_{\nd} \bfb y_{n}^{[\nd]} \, ,
\end{align*}
which can be then expressed in the form of \eqref{eq:theta:FG}, where 
$\bfb y_n = (\bfb y_n^{[0]}, \bfb y_n^{[1]}, \ldots, \bfb y_n^{[\nd]})$, and:
\begin{align*}
\bfb F &=
\begin{bmatrix}
\bfg 0_{rq, q} &  
\bfg I_{rq, rq} \\
\bfb M & -\bfb D \\
\end{bmatrix} , \ 
\bfb G =
\begin{bmatrix}
\bfg I_{rq,rq} & 
\bfg 0_{rq,q} \\
\bfg 0_{q,rq} & \bfb D_{\nd} \\
\end{bmatrix} , \\
\ \bfb D &= [\bfb D_0 \ \bfb D_1 \ \ldots \ \bfb D_{\nd-1}] \, .
\end{align*}
where $q = \nx+\ny$.

\end{proof}

\bibliographystyle{IEEEtran}
\bibliography{references}

\begin{biography}[{\includegraphics[width=1in,height=1.25in,clip,keepaspectratio]{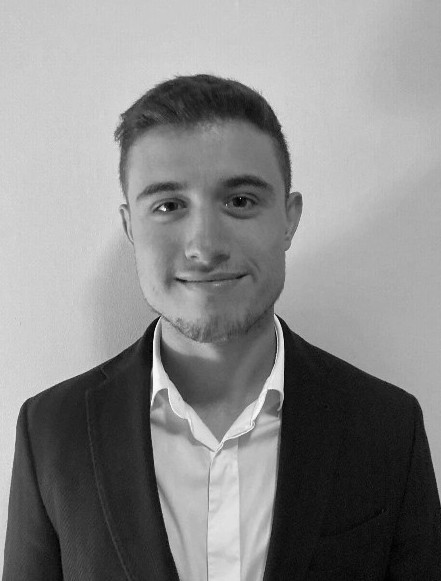}}]  {Andreas Bouterakos} 
received the Diploma (M.E.) in Electrical and Computer Engineering from the University of Patras, Greece, in 2021. Since September 2023, he has been a Ph.D. candidate with the School of Electrical and Electronic Engineering at University College Dublin, Ireland.  His current research interests include stability analysis and automatic control of power systems with high penetration
of distributed energy resources.
\end{biography}

\begin{biography}[{\includegraphics[width=1in,height=1.25in,clip,keepaspectratio]{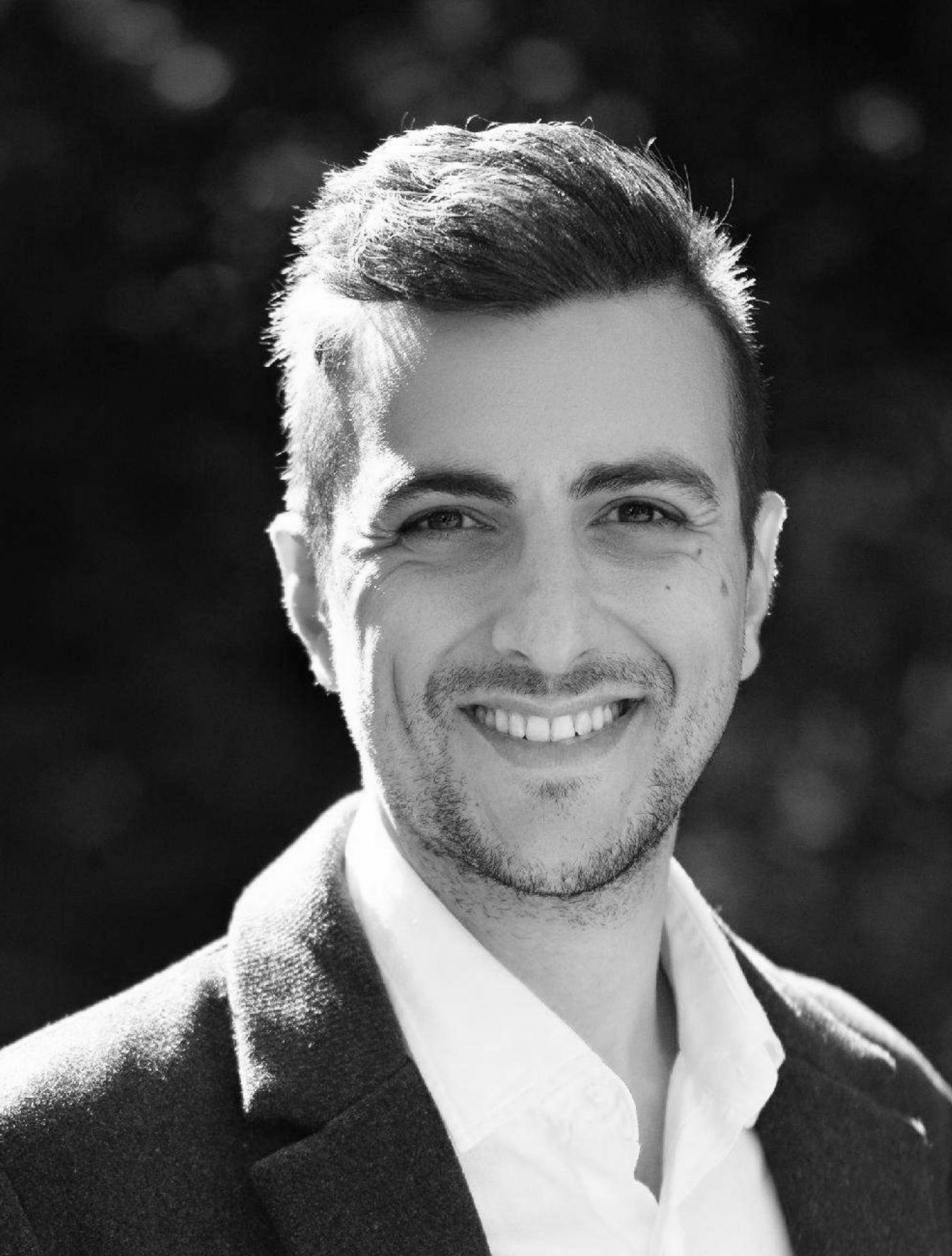}}]  {Georgios Tzounas} (M’21) received the Diploma (M.E.) in Electrical and Computer Engineering from the National Technical Univ.~of Athens, Greece, in 2017, and the Ph.D.~from University College Dublin (UCD), Ireland, in 2021.  In Jan.-Apr.~2020, he was a visiting researcher at Northeastern Univ., Boston, MA.  From Oct.~2020 to Apr.~2023, he was a postdoctoral researcher with UCD (2020-2022) and ETH Z\"urich (2022-2023).  Since Apr.~2023, he has been an Assistant Professor with the School of Electrical and Electronic Engineering at UCD.  His primary research area is power system dynamics.
\end{biography}

\vfill

\end{document}